\documentclass{article}

\usepackage{PRIMEarxiv}

\usepackage[utf8]{inputenc} 
\usepackage[T1]{fontenc}    
\usepackage{hyperref}       
\usepackage{url}            
\usepackage{booktabs}       
\usepackage{amsfonts}       
\usepackage{nicefrac}       
\usepackage{microtype}      
\usepackage{lipsum}
\usepackage{fancyhdr}       
\usepackage{graphicx}       
\graphicspath{{media/}}     
\usepackage{array} 
\usepackage{multirow}
\usepackage{amsmath,array}

\title{Investigation of the Influence of Macrozones in
Titanium Alloys on the Propagation and
Scattering of Ultrasound}

\author{
  Wei Yi Yeoh\thanks{Corresponding author: wyy18@ic.ac.uk} , Bo Lan , Michael J.S. Lowe\\
  Department of Mechanical Engineering\\\
  Imperial College London\\
}

\begin{document}
\maketitle

\begin{abstract}
The presence of macrozones (or micro-textured regions) in Ti-6Al-4V (Ti64) was shown to be a potential cause to the onset of cold dwell fatigue which reduces fatigue life significantly. Past research has demonstrated the potential of using ultrasonic testing for macrozone characterisation, with the variation of ultrasound attenuation, backscatter, and velocity in the presence of macrozones. However, due to the complexity of the microstructure, some physical phenomena that were observed are still not well understood. In this study, we propose the use of Finite Element (FE) polycrystalline models to provide us with a means to systematically study the wave-macrozone interaction. Through this investigation performed using two-dimensional (2D) models, we are able to identify important correlations between macrozone characteristics (size, shape, and texture) and ultrasound responses (attenuation, backscatter, and velocity). The observed behaviours are then validated experimentally, and we also highlight how this understanding can potentially aid with the characterisation of macrozones in Ti-64 samples.
\end{abstract}


\section{Introduction}
Ti-6Al-4V (Ti64) is a widely used alloy in the aerospace industry mainly because of its high strength-to-weight ratio and corrosion resistance \cite{Inagaki2014a,Peters2003}. It is mainly used in the “cold-section” of the aero-engine, such as the inlet fan blades and the compressor regions. The harsh operating conditions demand that the Ti64 components have superior mechanical properties to withstand the high cyclic loads and to delay the onset of mechanical fatigue with excellent fatigue resistance.

However, cold dwell fatigue (CDF) - a phenomenon in titanium alloys whereby stresses held at moderate temperatures can result in significant reductions in fatigue life \cite{Patel2018b} - is still not well-understood and thus cannot be incorporated reliably into the life predictions \cite{Dunne2008,Williams2018}. As such, conservative assumptions are used in both manufacturing and inspections to prevent the occurrence of CDF failure. CDF has been plaguing Ti-64 components and is believed to be the root cause for a number of catastrophic engine failures, such as the Southwest Airline incident in 2018 \cite{Southwest}.

\begin{figure}[h!]
	\includegraphics[width=60mm]{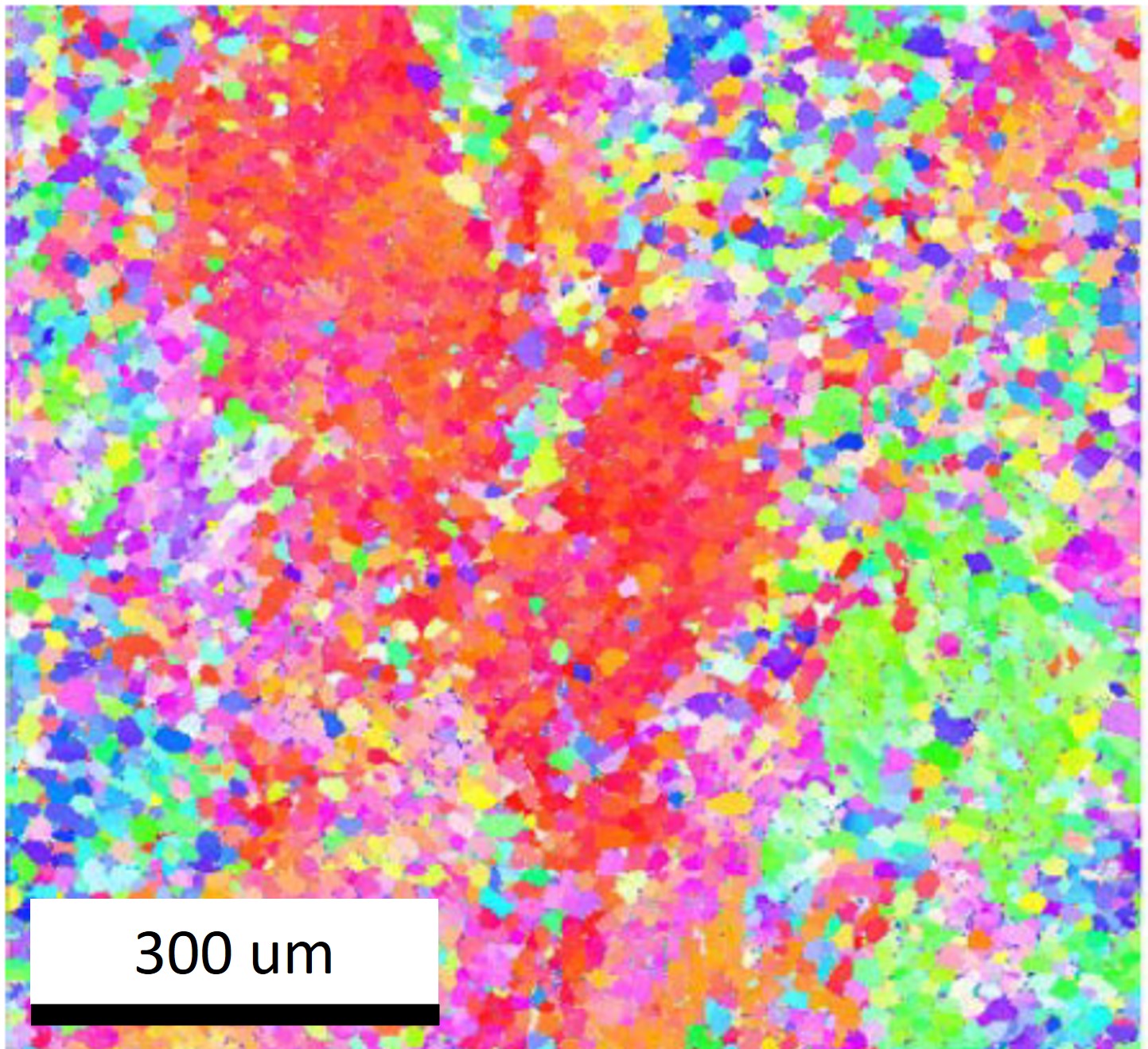}
	\centering
	\caption{EBSD image of a Ti-64 sample. The color code indicates specific grain orientations and the cluster of red in the EBSD image represents the presence of macrozones in the sample \cite{Bhattacharjee2011}. The key to the colour code is not given, because the significance here is the existence of regions of co-alignment of crystals rather than the values of the orientations.}
	\label{fig:Soft_Hard_Grain}
\end{figure}

Research over the past years uncovered that the presence of macrozones (also known as Macro-Textured Regions or MTRs) in Ti-64 is a potential cause to the onset of CDF \cite{Williams2018}. Macrozones refer to clusters of grains having similar preferential C-axis orientations which are retained from the growth of $\alpha$ colonies during the manufacturing process, as illustrated in Figure \ref{fig:Soft_Hard_Grain} \cite{Bhattacharjee2011}, with the color code corresponding to the orientation of the grains. C-axis is a term primarily used for hexagonal crystal structures to describe the direction along the height of the hexagonal prism or the [0001] axis. Past studies have shown that macrozone features such as the sizes, shapes \cite{Liu2021}, and C-axis orientations \cite{Zheng2016,Cuddihy2017,Xu2020} are potentially critical towards the onset of CDF - for instance, macrozones with high aspect ratios were shown to increase stress concentrations at the grain boundary by approximately 10\% \cite{Liu2021} which can be very damaging. Hence, we are primarily interested in evaluating these macrozone characteristics in this study.
Conventional inspection methods capable of evaluating the macrozones are generally destructive, time-consuming, and costly. For instance, metallography techniques such as Electron Backscatter Diffraction (EBSD) requires laborious sample preparations, which is destructive, and only provides local surface information \cite{Moreau2013,Liu2021}. Moreover, 2D images are not always representative of actual 3D characteristics. Hence, this prompts a real need to develop a non-destructive volumetric inspection method that can characterise the macrozones in Ti-64 efficiently.

Among the various non-destructive evaluation (NDE) techniques, ultrasonic testing shows great promise as it generally has low operating costs and contains superior penetration depth \cite{Hellier2003}. Moreover, the underlying elastodynamic principles cause key ultrasonic parameters such as attenuation, velocity, and backscatter to vary when employed on different materials. Extensive research in the past decades also demonstrated a good correlation between the ultrasonic parameters (attenuation, backscatter, and velocity) and microstructural features \cite{Matori2015}. Ultrasound attenuation was shown to be correlated to grain size over three distinct frequency regimes, namely the Rayleigh, Stochastic, and Geometric regimes \cite{Stanke1984a,Weaver1990,VanPamel2018a}. This attenuation-relation was further developed within the Stochastic regime to include the effects of elongated grains \cite{Yang2011c,Huang2021}. Next, ultrasound backscatter measures the scattered signals generated from grains and was shown to be related to grain sizes \cite{Rose1991,Margetan1991,Yu2010,Liu2019} and was subsequently modified to account for variations in grain shapes \cite{Li2014a,Arguelles2016b}. Lastly, ultrasound velocity can be used to determine grain orientations \cite{Smith2014,Patel2018b} and bulk material texture \cite{Lan2014a,Bo2018}. 

Specific to macrozone characterisation using ultrasound, Bhattacharjee \cite{Bhattacharjee2011} and Margetan \cite{Margetan2005} showed positive correlation between attenuation and the average macrozone sizes experimentally with normal incidence ultrasound scans. Blackshire et al. demonstrated experimentally that macrozones can generate coherent backscatter signals and validated his findings with wave-propagation finite-element models in a polycrystalline model incorporated with actual EBSD data \cite{Blackshire2018,Blackshire2019}. Rokhlin et al. were able to invert bulk macrozone sizes using backscatter ratios which showed good agreement with EBSD data \cite{Yang2012c,Pilchak2014b}. This was further refined by incorporating attenuation measurements to invert elastic scattering factors of the bulk material \cite{Rokhlin2021a}.

We seek to expand this understanding by evaluating the propagation and scattering of ultrasound in a range of Ti-64 material samples with varying macrozone sizes, shapes, and preferred crystal orientations. The bulk of the theoretical solutions rely on simplifications and in some cases on statistical methods. However, actual microstructures are difficult to control and can only validate parts of the theory and at times may also lack a detailed description of the wave field interactions. Computational models, on the other hand, can generate perfectly-controlled models to provide physical insights to the observed phenomena and to validate and extend the existing theoretical solutions. To achieve this, we make use of Finite Element (FE) simulations which serves as perfectly-controlled experiments since microstructural features in actual Ti-64 samples are difficult to control. This exploits the recent validated capability to simulate ultrasound wave propagation through polycrystals accurately, with grain scale representation \cite{VanPamel2018a,Huang2021,huang2020}. Therefore, we see the value in an overview study that demonstrates the relative sensitivity of ultrasound to different macrozone parameters and to identify what may be most useful for future detailed developments. Apart from delving into empirical investigation, we also aim to uncover the physical insights for the observed behaviours to corroborate with the extensive work conducted in the past. This enables us to highlight important correlations that can potentially be further developed and deployed into a quantitative inversion measurement method capable of characterising macrozone sizes in Ti-64.

In section 2, the approach and methodology used in this study is described. In section 3, the results obtained from the finite element models are detailed and discussed. In section 4, we evaluate the FE results with experimental validations. Lastly, in section 5, a conclusion and potential future work are provided.

\section{Methodology}
\label{sec:Methodology}

\subsection{Finite Element Model Configuration}
In this FE study of wave propagation in polycrystals, the goal is to understand the physical interactions between ultrasound wave propagation and macrozones of various characteristics. Hence, 2D models are used as opposed to 3D to increase computational efficiency. Even though it is well documented that 2D and 3D models have different quantitative behaviours in attenuation and backscatter - attenuation for instance, was shown to have a dimensional dependence when operating in the Rayleigh regime \cite{VanPamel2016i,Bai2020} - the physical insights are still transferable we believe that they are qualitatively representative of actual 3D behaviour. This is sufficient for the present study whose aim is to identify the nature of the behaviour and trends.

The model setup is shown in Figure \ref{fig:FE_Model_Setup} which comprises polycrystalline microstructure. The grains were generated using the Neper software \cite{Quey2011} and the Voronoi tessellation method \cite{Quey2011} with an average grain size of 20 $\mu$m \cite{Liu2021}. The model has a domain of 20 by 20 mm and contains 1,150,000 individual grains. Crystal orientations for the grains are specified using Roe's formulation and passive rotations \cite{Lan2015d} with the following stiffness matrix (unit: GPA) for the Ti-64 alpha phase \cite{Lan2015c}:

\begin{equation} \label{EC}
    C = 
    \begin{pmatrix}
        170 & 92 & 70 \\
        92 & 170 & 70 \\
        70 & 70 & 192 \\
        & & & 52 \\
        & & & & 52 \\
        & & & & & 39
    \end{pmatrix}
\end{equation}
Ti-64 alpha phase has a hexagonal crystal structure which is isotropic about its C-axis. The velocity profile in the 2D X-Y plane can be generated by solving the Christoffel equation using the stiffness matrix and is illustrated in Fig 2, whereby velocity is highest along the C-axis and lowest normal to the C-axis.

\begin{figure}[h!]
	\includegraphics[width=140mm]{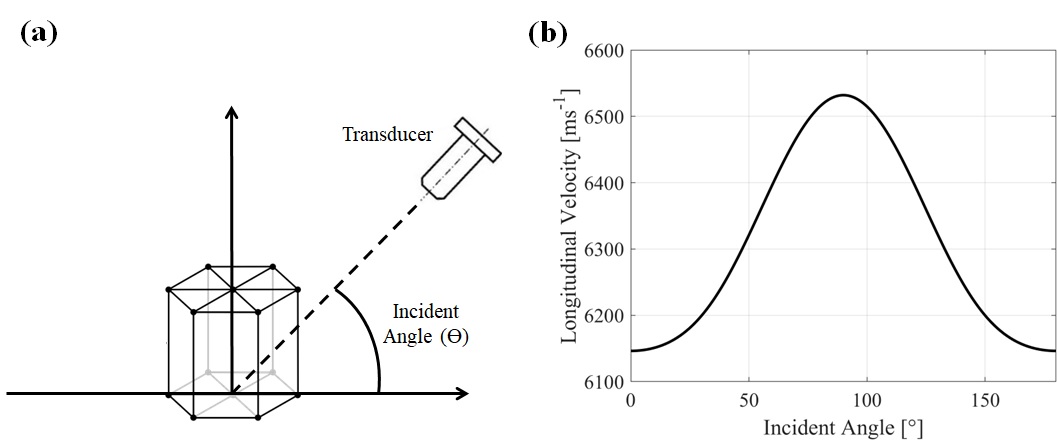}
	\centering
	\caption{(a) A hexagonal crystal structure with the incident angle defined and (b) the corresponding longitudinal wave velocity with varying incident angle.}
	\label{fig:FE_Model_Setup}
\end{figure}
The grains are given randomised orientations to generate an untextured medium with a Voigt-averaged velocity of 6244 m/s. Subsequently, regions of interest are identified and the grains within them are given the same crystal orientation to form larger grains that are representative of macrozones - this is an assumption that we believe is reasonable for the study, as grain orientations in actual macrozones contain certain statistical preference \cite{Venkatesh2020}.
A periodic arrangement of the macrozones with an arbitrary density was used in this study for simplification such that a comparative analysis can be made between models with variations in the macrozone characteristics and to reflect a more realistic scenario since actual samples usually contain multiple macrozones. Throughout this study, we also made use of single macrozone models to simplify and explain several physical effects that arise in the wave-macrozone interactions which is present in the multi-macrozone models in a more complex manner.
Based on the numerical studies reported in \cite{VanPamel2015f}, it is recommended that the mesh size used be smaller than one-twentieth of the simulated wavelength and one-tenth of the average grain diameter to obtain accurate simulation models. An excitation frequency of 10 MHz is used, and the Voigt-averaged longitudinal velocity is 6244 m/s. This gives a wavelength of approximately 0.6 mm at the centre frequency, and a range of 0.5 to 1 mm over the excitation bandwidth. Using the aforementioned criteria, the model is spatially discretized using a structured mesh of 2 $\mu$m square elements. Both structured and unstructured meshes were shown in \cite{VanPamel2017a} to offer similar performance with the appropriate discretisation size. Hence, for the sake of simplicity, structured mesh was selected for this application.

We use an explicit time domain computation coupled with a central difference time marching scheme which incrementally solves the system of equations with a time step of 0.1 ns to satisfy the Courant-Friedrichs-Lewy (CFL) condition \cite{VanPamel2018a}. To excited a longitudinal wave, a 3-cycle Hann-windowed cosine toneburst force excitation is applied to the nodes located at the top surface of the model in the Y-direction, see Figure \ref{fig:FE_Model_Setup}a. To support the propagation of the plane wave, symmetrical boundary conditions are applied to the outer left and right surfaces of the model by constraining the X-direction displacements. This simulates an infinitely wide medium which is ideal for isotropic materials. In the case of polycrystals, symmetrical boundary conditions can only approximate a plane wave which was shown to be a reasonable assumption in \cite{huang2020}. The FEA model is then computed using Pogo, a GPU-based finite element solver which can accommodate larger models and provide faster solutions than the conventional CPU-based solver \cite{Huthwaite2014}. The FE simulation of elastic wave propagation in polycrystalline materials have also been extensively tested in \cite{huang2020} to maximise the accuracy of the results obtained.

\begin{figure}[h!]
	\includegraphics[width=130mm]{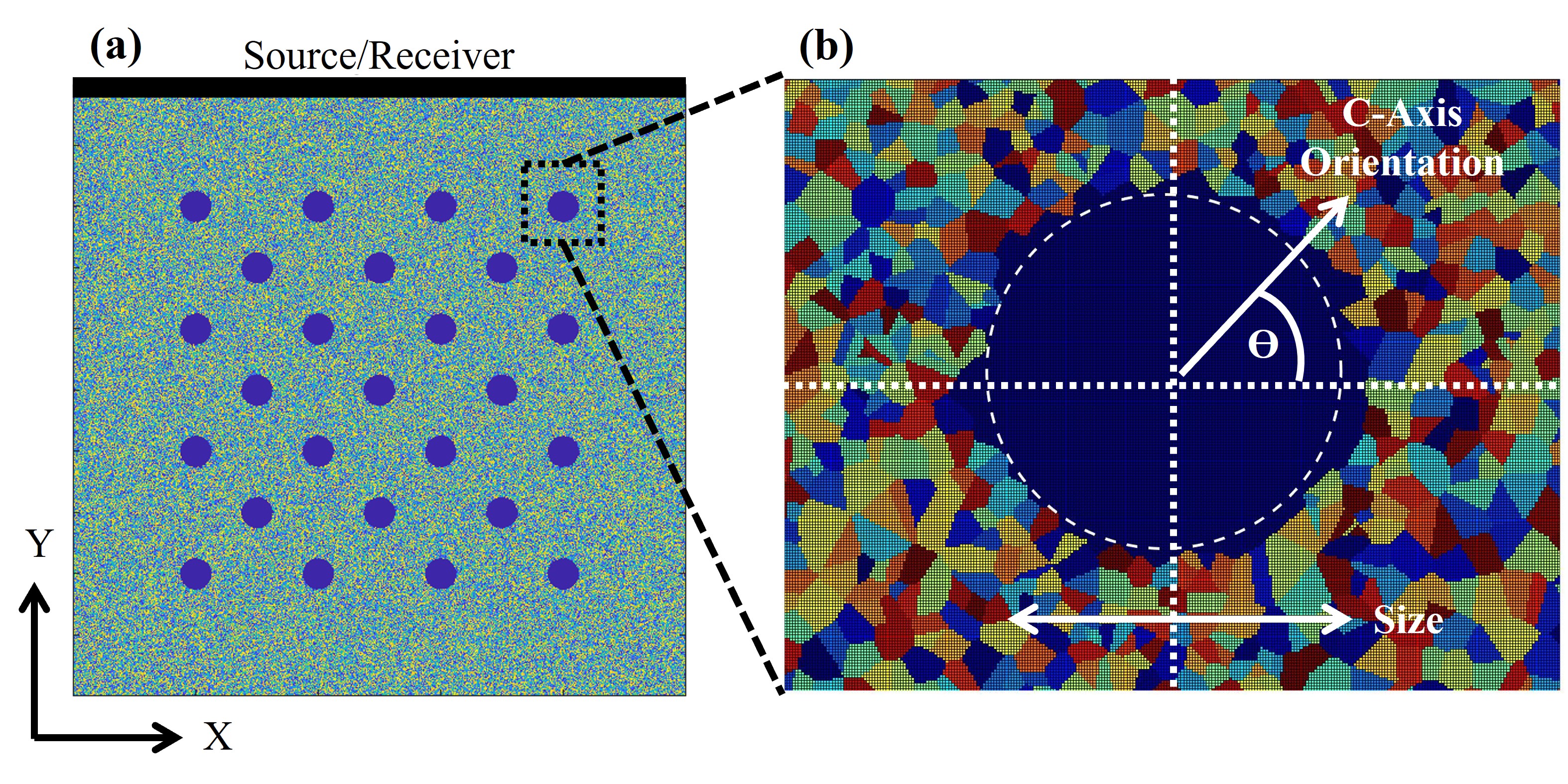}
	\centering
	\caption{(a) 2D FE model setup with multiple macrozones (blue) located in the medium in a background filled with polycrystals. The top surface (in black) is the location for the excitation and monitoring nodes. (b) A zoomed-in image of a single macrozone surrounded with individual polycrystals of random orientations as depicted by the color codes, with the size and C-axis orientation of the macrozones with respect to the wave propagation direction labelled.}
	\label{fig:FE_Model_Setup}
\end{figure}

\begin{figure}[h!]
	\includegraphics[width=90mm]{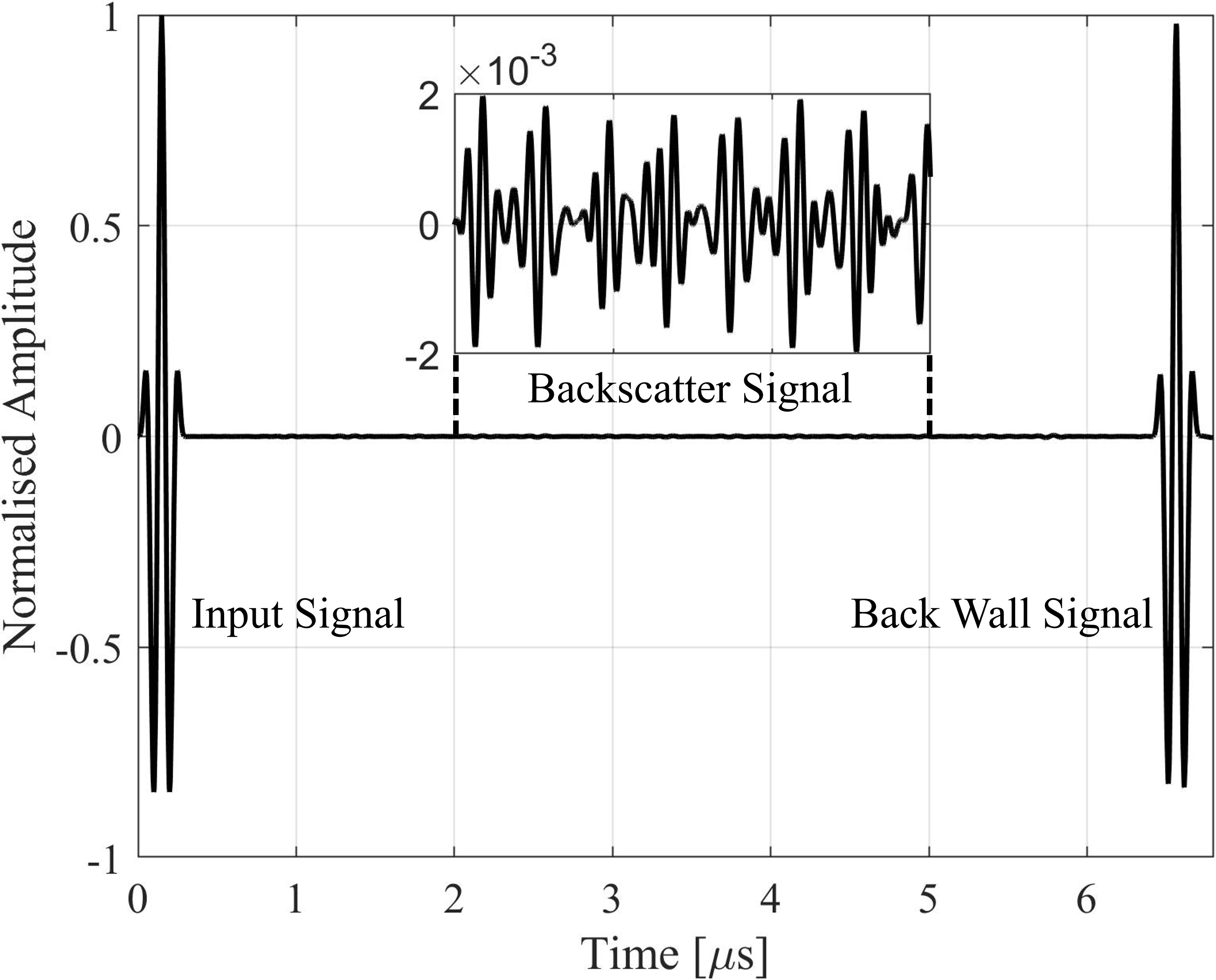}
	\centering
	\caption{FE simulated time domain signal which comprises the front wall (input) and back wall signal. The inserted figure represents the backscatter signal which is presented on a much smaller amplitude scale.}
	\label{fig:FE_Signal}
\end{figure}

Once the FE models are solved, the time-domain signals recorded at the monitoring nodes are extracted for post-processing. The excited wave propagates from the top down to the bottom and undergoes reflection, to be received back at the top. The signals are then summed across all the receiving nodes. An example is shown in Figure \ref{fig:FE_Signal} where the input signal (front wall, top), the backscattered, and the signal reflected from the back surface (back wall, bottom) of the model can be seen. These features are isolated by applying a tapered window so that they can be used for attenuation, backscatter, and velocity measurements.

For general ultrasound attenuation measurements, we take the amplitude ratio between the front and back wall signals. This is normally compensated by accounting for the interface transmission and reflection coefficients, and a beam-diffraction compensation factor to account for beam spread when using a finite-sized transducer \cite{Rogers1974a}. This is represented by the formula \cite{Zeng2010}:
\begin{equation} \label{att}
\alpha_s = \frac{1}{z} \ln\left( \frac{FD_1 T_{01}R_{10}T_{10}}{BD_0 R_{01}} \right)
\end{equation}
where $z$ is the propagation distance, $F$ and $B$ are the front and back wall signal amplitude in the frequency domain respectively, $D$ is the beam diffraction coefficient, $T$ and $R$ are the transmission and reflection coefficients respectively, and the indices 0 and 1 represents water and Ti-64 medium respectively. Since our model comprises a single solid medium with plane wave propagation, the diffraction correction factor and the reflection/transmission coefficients are neglected. The isolated front and back wall signals are Fourier Transformed to obtain their respective frequency contributions and subsequently applied into equation \ref{att} to obtain an attenuation profile with respect to frequency. 

Next, for backscatter signals, a common processing method is to generate a backscatter amplitude profile by conducting a root mean square (RMS) operation across multiple spatial measurements \cite{Margetan1993,Margetan1994}. However, in this parametric study, only a single time-domain signal is obtained for each macrozone model configuration. Hence, to provide a convenient measure of the overall backscatter response such that it can be used to compare between the models, the backscatter intensity is calculated. The backscatter signals are isolated in the time domain from 1 to 6 us and Fourier Transformed into the frequency domain. Next, the magnitudes of the backscatter spectral response are summed within the full width half maximum bandwidth of the excited signal to produce a single backscatter intensity value.

Lastly, for ultrasound velocity measurements, the cross-correlation method is used. Cross-correlation measures the similarity between two signals as one is shifted in time \cite{Zhang2008}. This is represented mathematically by an inner product in the time domain:

\begin{equation} \label{CrossCorrelation}
R_{xy}(t) = \int_{-\infty}^{\infty} X(\tau)Y(\tau - t) d\tau
\end{equation}

where $R_{xy}$ is the cross-correlation function, $X$ and $Y$ are the two signals to be correlated, and $t$ is the delay interval. Using cross-correlation, a delay interval from the isolated front and back wall signals is generated, and dividing this time-of-flight (TOF) by the propagation distance of 20 mm gives the wave velocity.

\subsection{Computational Analysis Approach}
To better understand the physical phenomena that arise from the wave propagating through these complex macrozone features, and to identify how strongly the respective factors are affecting the propagation, an empirical approach is considered in this paper. This is conducted using Design of Experiments (DOE) or experimental design, which is a convenient and systematic tool that allows us to identify the relative sensitivity of multiple parameters to various responses with a series of models or tests \cite{Montgomery2017}. The statistical software Minitab \cite{Minitab} uses the Analysis of Variance (ANOVA) method, followed by a F-test to determine the significance of the factors and their impact on the responses with a predefined confidence interval (95\%) \cite{Montgomery2017}. This is only valid within the limits of the modelled parameters.

To effectively evaluate the relative sensitivity of the different macrozone characteristics, we generate multiple models with different macrozone shapes, sizes, and crystal orientation as listed in Table \ref{fig:Macrozone_Parameters}. A total of 125 FE models are generated where the listed parameters are varied individually. These FE models are generated following the procedure described in section 2(a). The time domain signals are captured to evaluate the ultrasound attenuation, backscatter, and velocity responses. The definition of the characteristics is illustrated in Figure \ref{fig:FE_Model_Setup}(b). The size refers to the spherical diameter, varying from 0.1 mm to 1 mm which covers a range of possible macrozone sizes \cite{Liu2021}. The shape of the macrozones is termed with the alphabet representing the axis of elongation, and the number representing the elongation ratio (for example, X4 - elongated 4:1 in X-direction, XY1 - equiaxial). The total area occupied by the equiaxial and elongated macrozones is held constant. The C-axis orientation of the macrozones is defined with respect to the normal of the wave propagation direction. The randomly oriented grains in the background have a mean diameter of 20 um.

\begin{table}[!h]
\begin{center}
\caption{Macrozone characteristics used in the parametric study.}
\label{fig:Macrozone_Parameters}
\begin{tabular}{|>{\centering\arraybackslash}m{3cm}| >{\centering\arraybackslash}m{3cm}|>{\centering\arraybackslash}m{5cm}|}
\hline
Size [mm] &Shape &MTR Crystal Orientation [$^{\circ}$] \\
\hline
0.1 &X4 &0 \\
0.25 &X2 &30 \\
0.5 &XY1 &45 \\
0.75 &Y2 &60 \\
1 &Y4 &90 \\\hline
\end{tabular}
\vspace*{-4pt}
\end{center}
\end{table}

The ultrasound responses are then used as inputs for the Minitab software \cite{Minitab} which generates statistical information describing the correlation between macrozone characteristics (factors) and the responses. The steps can be represented by the following equations to calculate the Sum of Squares (SS),  Sum of Squares Error (SSE), Mean Squares (MS), Mean Squares Error (MSE), and the F-ratio (F). SS measures the variation attributed to the varying macrozone characteristics, whereas SSE represents the variation attributed to error or the difference between the individual measurements and the predicted values. They are represented by:

\begin{equation} \label{SS}
SS = \sum (\Hat{X} - \Bar{X})^2 
\end{equation}

\begin{equation} \label{SS}
SSE = \sum (X - \Hat{X})^2 
\end{equation}

where X is the measured value, $\Bar{X}$ is the mean value, and $\Hat{X}$ is the predicted value. MS is used to estimate the variance between the sample means and MSE represents the variation within the sample means. They are calculated by taking the following ratios:

\begin{equation} \label{MS}
MS = \frac{SS}{DOF}
\end{equation}

\begin{equation} \label{MS}
MSE = \frac{SSE}{DOF_{Err}}
\end{equation}

where DOF is governed by the number of data points available (sample size) and the number of terms that are being estimated. $DOF_{Err}$ are the remaining points used to estimate the errors within the model. The F-ratio is calculated by taking the ratio of the MS and the Mean Squares Error (MSE):
\begin{equation} \label{FV}
F = \frac{MS}{MSE}
\end{equation}

which indicates the significance of each factor present in the experimental design. The results are then presented in the form of a Pareto chart which indicates the order of significance of the factors, and a main effects plot which shows the trend and relationship between the factors and the responses \cite{Montgomery2017}.

\section{Finite Element Modeling Results}
This section describes the FE results to better understand why certain mechanisms occur and how they can be useful in providing some form of qualitative inversion towards the microstructure of the system. We will evaluate the Pareto Chart and the Main Effects Plot generated using Minitab and identify the relative significance of each macrozone characteristic and their correlation to the respective ultrasound responses.

\subsection{Attenuation}
\begin{figure}[h!]
	\includegraphics[width=80mm]{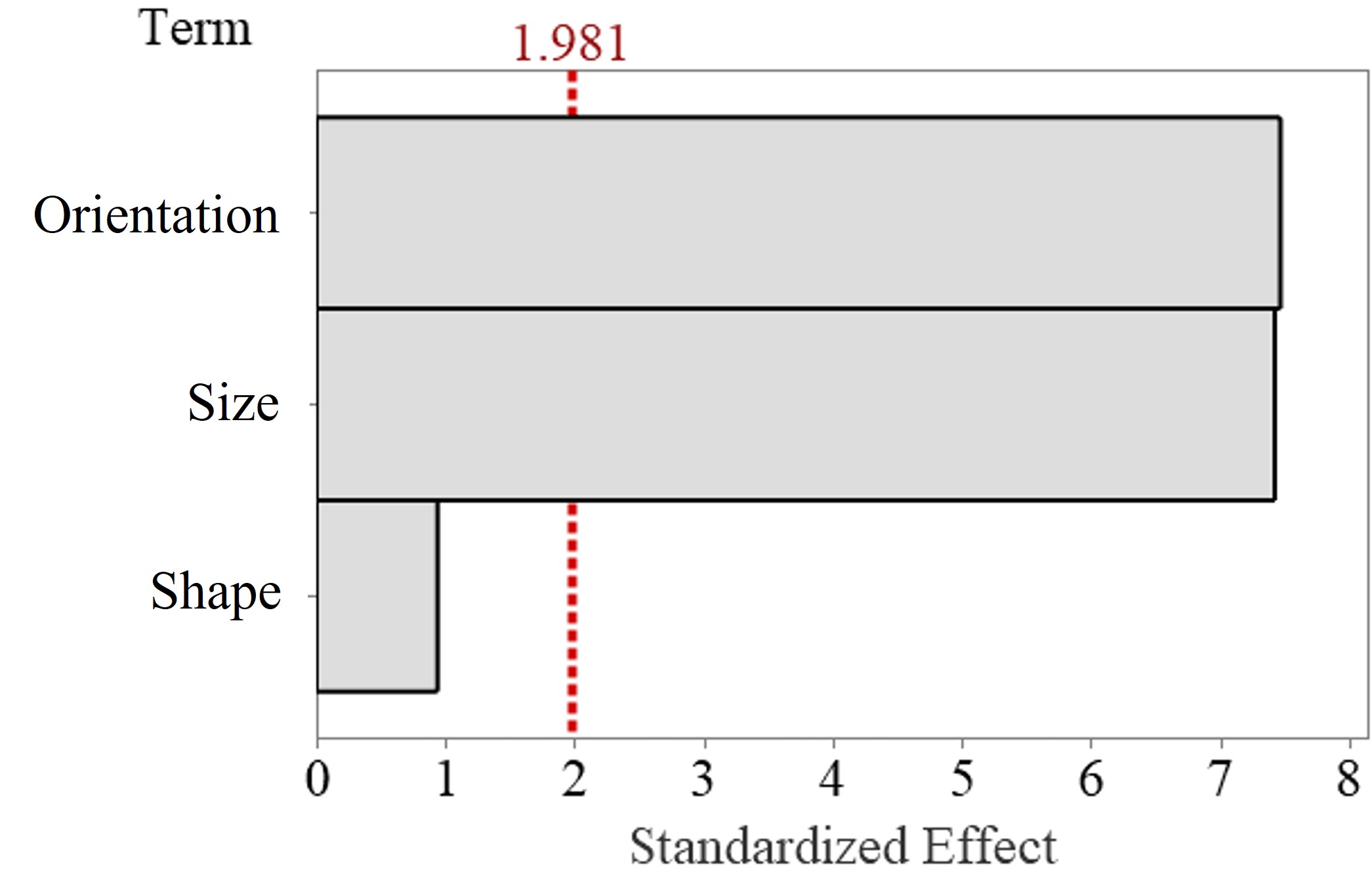}
	\centering
	\caption{Pareto chart of the standardized effects for attenuation predicted at a 95\% confidence interval. The red line (1.981) indicates that factors exceeding this are deemed to have a statistically significant impact.}
	\label{fig:Pareto_Att}
\end{figure}

We first evaluate how ultrasound attenuation varies with different macrozone characteristics. From the Pareto Chart in Figure \ref{fig:Pareto_Att}, the different macrozone characteristics are labelled on the Y-axis, with the standardized effect on the X-axis. The standardized effect is a normalised scale that represents the significance of the factors, with bars exceeding the red-dotted line being deemed as statistically significant - the line is drawn at the (1-$\alpha$/2) quantile of a t-distribution, where $\alpha$ is the level of significance \cite{Montgomery2017}. From the results, macrozone crystal orientation has the greatest impact, followed by macrozone size, and lastly the macrozone shape which is deemed to be statistically insignificant within our range of modelled parameters.

\begin{figure}[h!]
	\includegraphics[width=130mm]{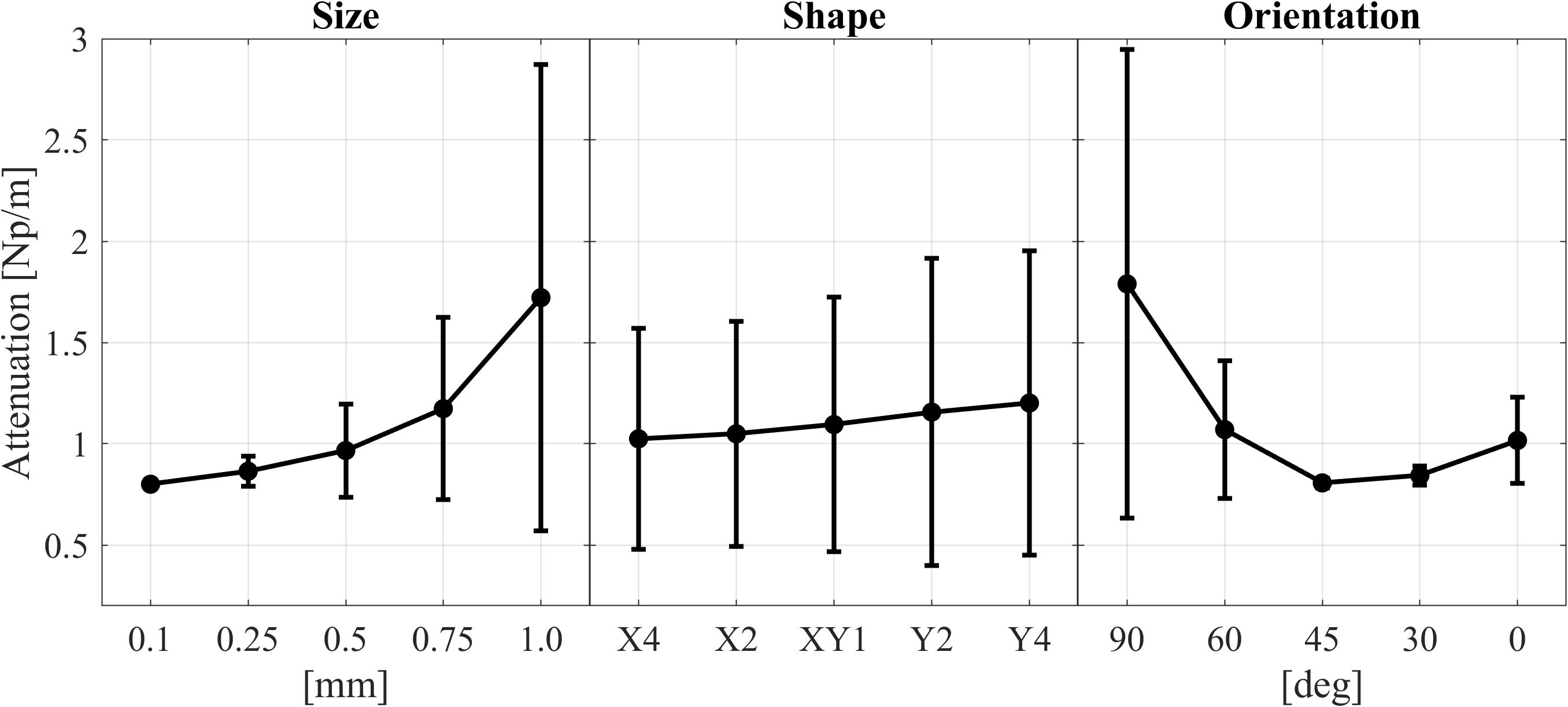}
	\centering
	\caption{The main effects plot for attenuation highlights the correlation between ultrasound attenuation and the three macrozone characteristics: (a) size, (b) shape, and (c) orientation.}
	\label{fig:Main_Effects_Att}
\end{figure}

The Pareto chart is accompanied by the main effects plot which shows the trends of the various macrozone characteristics with respect to attenuation. The main effects plot is illustrated in Figure \ref{fig:Main_Effects_Att}, with the dots and error bars on the plot representing the mean and standard deviation respectively of the attenuation computed across the models that contains the macrozone characteristic labelled in the X-axis. Several analyses can be made based on the mean values. An increase in macrozone dimension - which is dependent on the size and shape - along the wave propagation direction leads to an increase in attenuation. Crystal orientation affects the velocity mismatch between the macrozones and the surrounding materials and thus also results in varying attenuation values.

\begin{figure}[h!]
	\includegraphics[width=110mm]{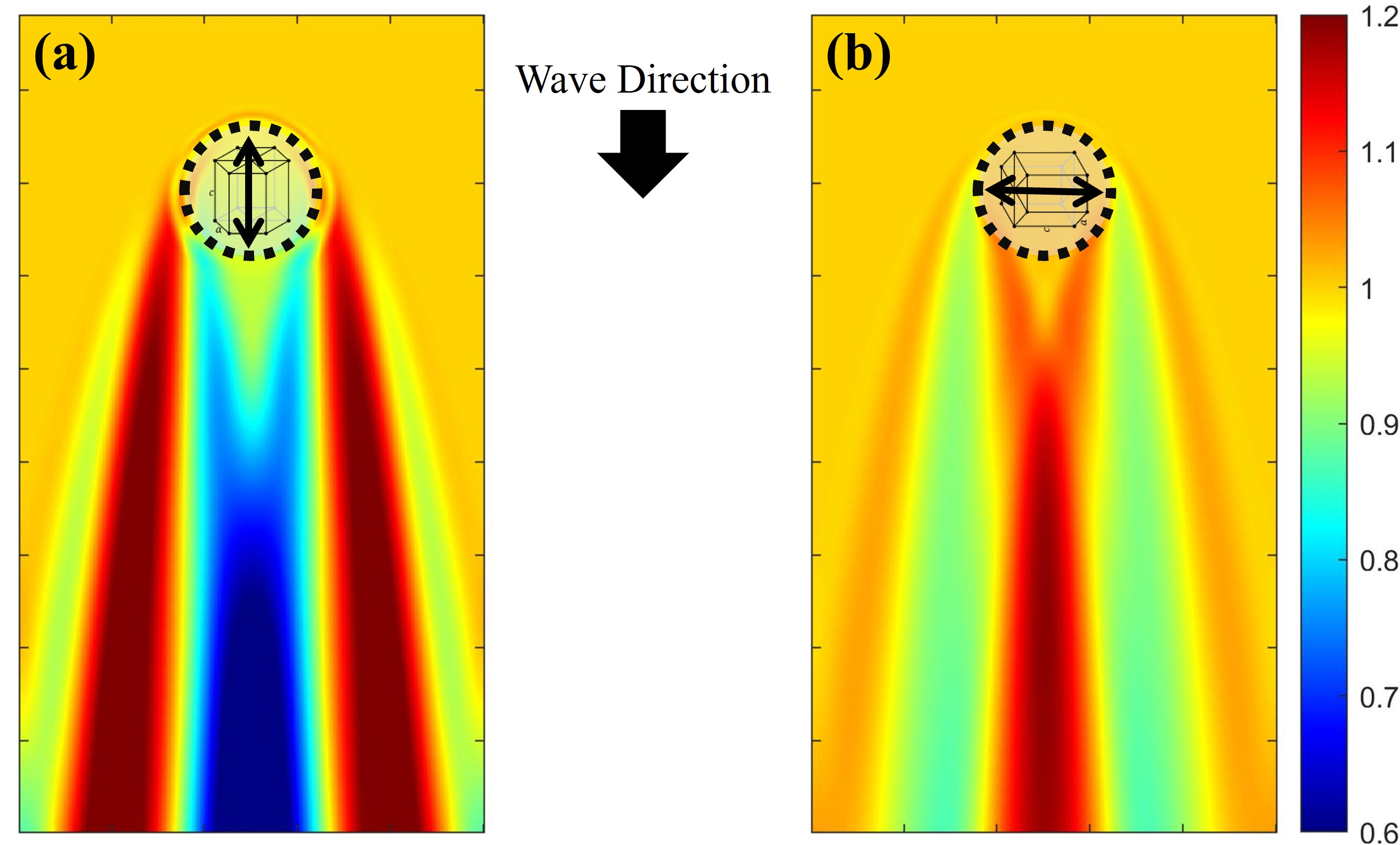}
	\centering
	\caption{Beam profiles generated in 2D FE models by taking the maximum amplitude at each spatial node within the model, where the color code represents the normalized arbitrary wave amplitude. The profiles show the distortion of the plane longitudinal wave after it interacts with a single circular macrozone of (a) parallel and (b) normal C-axis crystal orientation, generating a defocusing and focusing effect respectively.}
	\label{fig:Beam_Distortion}
\end{figure}

It is a well-known fact that any increment of grain sizes (or grain-size to wavelength ratio) leads to an increase in attenuation due to the larger grains and boundaries that scatters and distorts the propagating wave \cite{Stanke1984a}. This is the case for macrozones as well, but with the additional preferred crystal orientation that exisits within the macrozone as compared to the untextured polcrystalline background. This can be represented by a beam profile shown in Figure \ref{fig:Beam_Distortion}. The beam profile is plotted by taking the maximum displacement at each spatial node within the model domain throughout the time period of the simulation. The color code indicates the normalised displacement amplitude which varies after the wave interacts with the macrozone. On the left, the model contains a single macrozone embedded in a Voigt-averaged homogeneous medium. The macrozone C-axis is parallel to the wave propagation direction which travels from the top to the bottom. Wave velocity along the Ti-64 hexagonal crystal C-axis is higher at about 6550 m/s and thus the macrozone becomes a region of higher velocity. This leads to a local focusing effect which increases the wave amplitude across the wake of the macrozone, and vice versa. This results in a phase perturbation of the wave front, causing destructive interference to occur when the whole beam is captured by a receiver and this is represented as attenuation. Attenuation is greater along the C-axis orientation (90 $\deg$) because of the higher velocity mismatch with the Voigt-averaged background velocity of 6244 m/s, leading to a stronger distortion effect. With increased macrozone sizes, the larger grain boundary increases the scattering effects \cite{Zeng2010} while the larger grain area increases the beam distortion effect, both of which contribute to higher attenuation.

\begin{figure}[h!]
	\includegraphics[width=80mm]{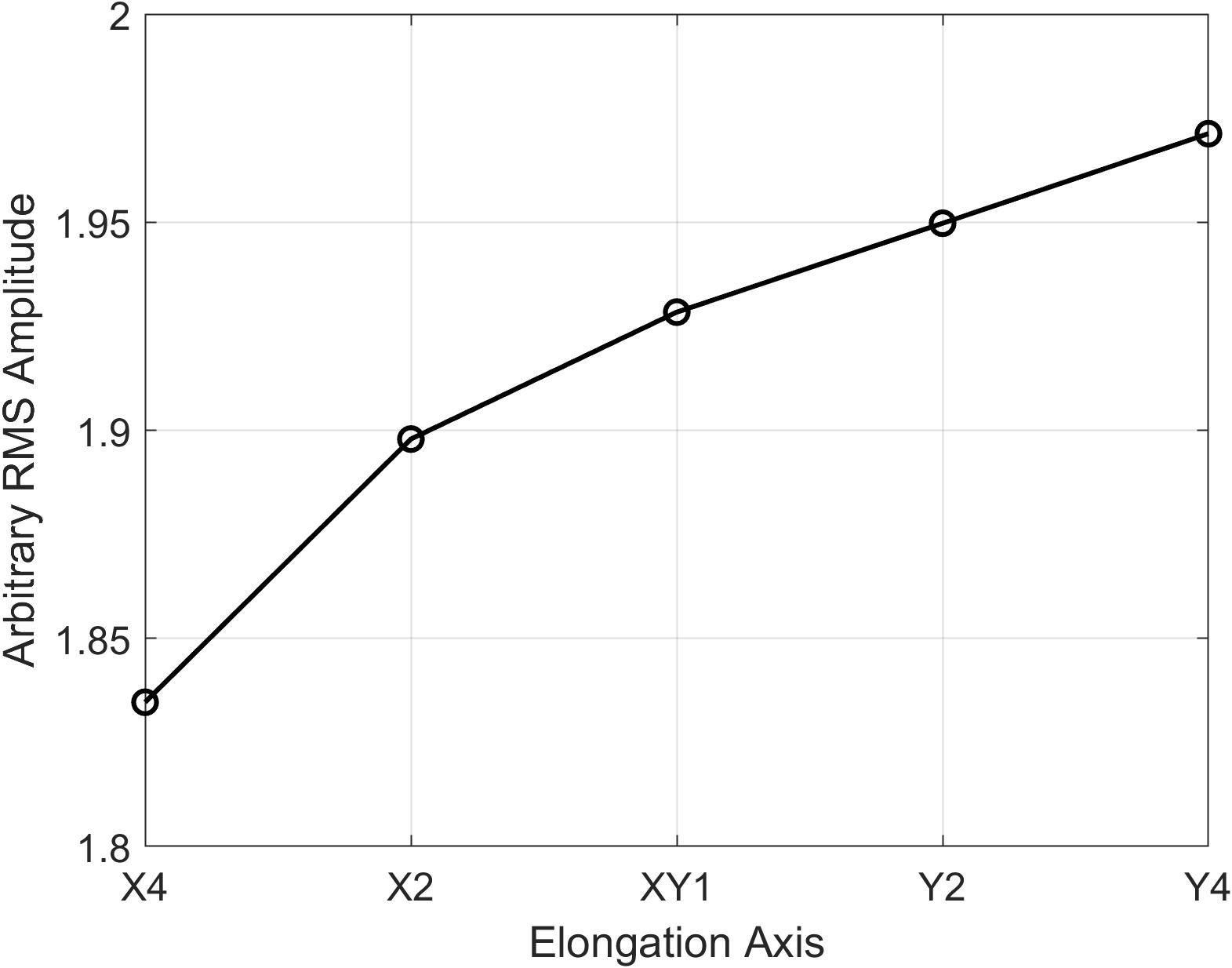}
	\centering
	\caption{RMS values calculated from the individual nodal displacement values across the receiver located at the bottom of the model for macrozones of different shapes, indicating a positive correlation between macrozone-length along wave propagation direction and attenuation.}
	\label{fig:Displacement}
\end{figure}

With changes in macrozone shapes from X4 to Y4, the macrozone-length along the wave propagation direction increases. Past study has shown that grain elongation within the Stochastic regime (wavelength being comparable to grain size) results in increased attenuation due to greater phase perturbation \cite{Ming2020}. Even though our models only have a limited number of macrozones embedded in an equiaxed polcrystalline medium, the phase perturbation effects due to macrozone elongation are also present. To quantify this perturbation effect, an RMS value representing the wave displacement fluctuations is calculated across the receiving nodes for each of the models. The displacements across the receiving nodes are normalised by the spatially averaged displacement to have a mean value of 0. Hence, the RMS values calculated from the normalised displacement correlates to the wave field fluctuations \cite{Ming2020}. As seen in Figure \ref{fig:Displacement}, the RMS value increases from 1.835 to 1.971 when the grain-length increases by a factor of 16 (from X4 to Y4). This indicates that macrozones with longer grain-length along the wave propagation direction results in stronger local distortions and thus a higher measured attenuation.

\begin{figure}[h!]
	\includegraphics[width=130mm]{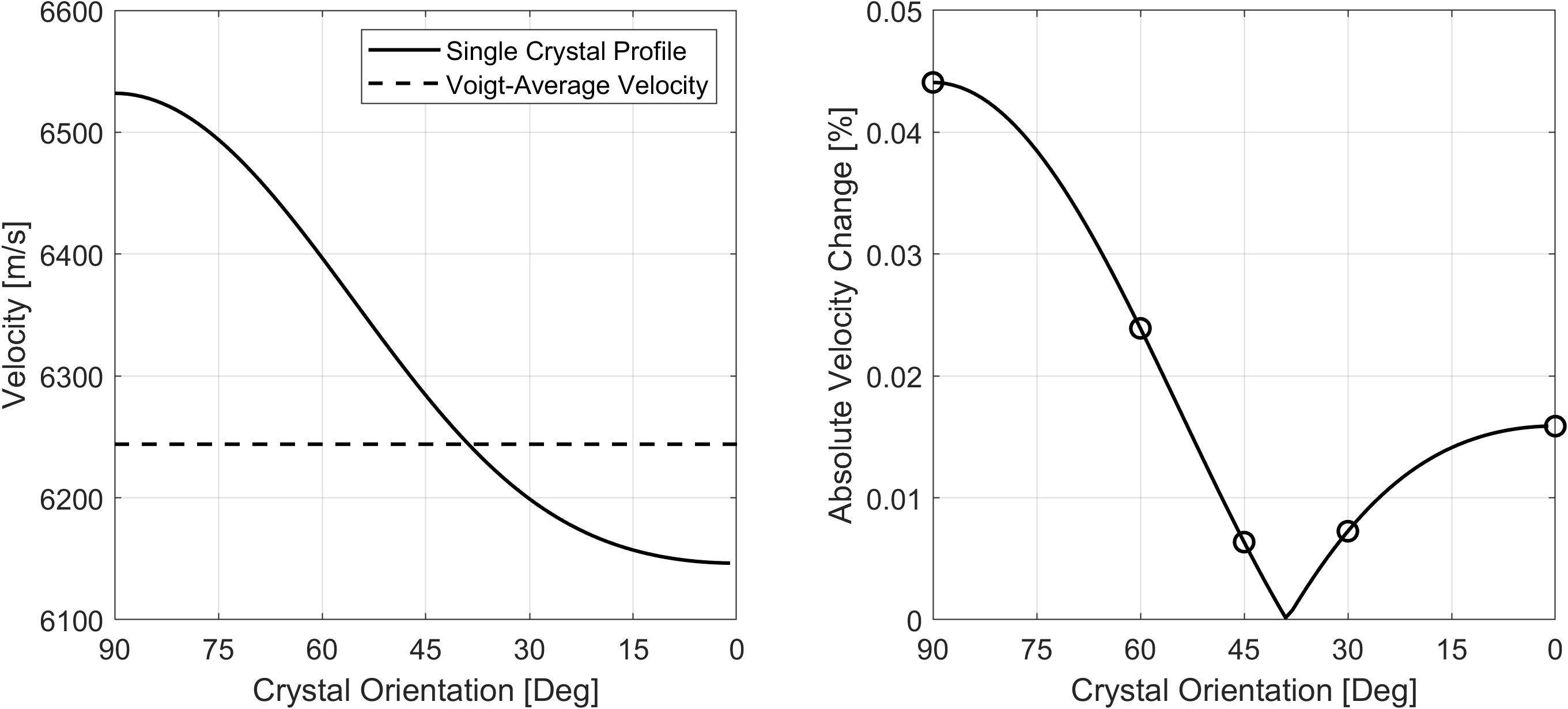}
	\centering
	\caption{(a) Ti-64 single crystal velocity profile with different C-axis orientation and the Voigt-averaged velocity value, and (b) Velocity contrast with different C-axis orientation with respect to Voigt-averaged velocity.}
	\label{fig:Velocity_Profile}
\end{figure}

Lastly, for the case of the macrozone crystal orientation, the attenuation profile has a minimum point at about 45 degrees as shown in Figure \ref{fig:Main_Effects_Att}. This is linked to the velocity mismatch between the macrozones and the untextured polycrystalline background in the model \cite{Lan2014a}. The velocity profile of a Ti-64 single crystal is illustrated in Figure \ref{fig:Velocity_Profile} (a) in blue, with the Voigt-averaged velocity shown in red. In Figure \ref{fig:Velocity_Profile} (b), the velocity difference between the single-crystal and the Voigt-averaged case is shown, with a minimum point at 39 degrees and has a similar profile to the main effects plot shown in Figure \ref{fig:Main_Effects_Att}. This velocity difference is positively correlated to ultrasound impedance which leads to the effects of wave scattering.

Our simulation model is populated with a finite number of macrozones with a predefined crystal orientation that introduces the presence of texture that affects the Orientation Distribution Function (ODF). ODF is a measure of the overall orientation of the grains in a region of interest and is related to bulk texture. Liu et al. \cite{Li2015b} demonstrated analytically that macroscopic texture in a polycrystalline material generally results in a lower attenuation as compared to macroscopic isotropy due to the reduced crystal misorientations between neighbouring grains. In our model, apart from macroscopic texture, the location of the macrozones that are defined periodically also changes the Misorientation Distribution Function (MDF) – MDF measures the level of crystal misorientation between neighing grains and contains additional spatial information. Both ODF and MDF influences the attenuation of ultrasound through scattering and wave distortions and as such the simulated attenuation in our models are generally higher than the macroscopically isotropic case.

\subsection{Backscatter}
\begin{figure}[h!]
	\includegraphics[width=80mm]{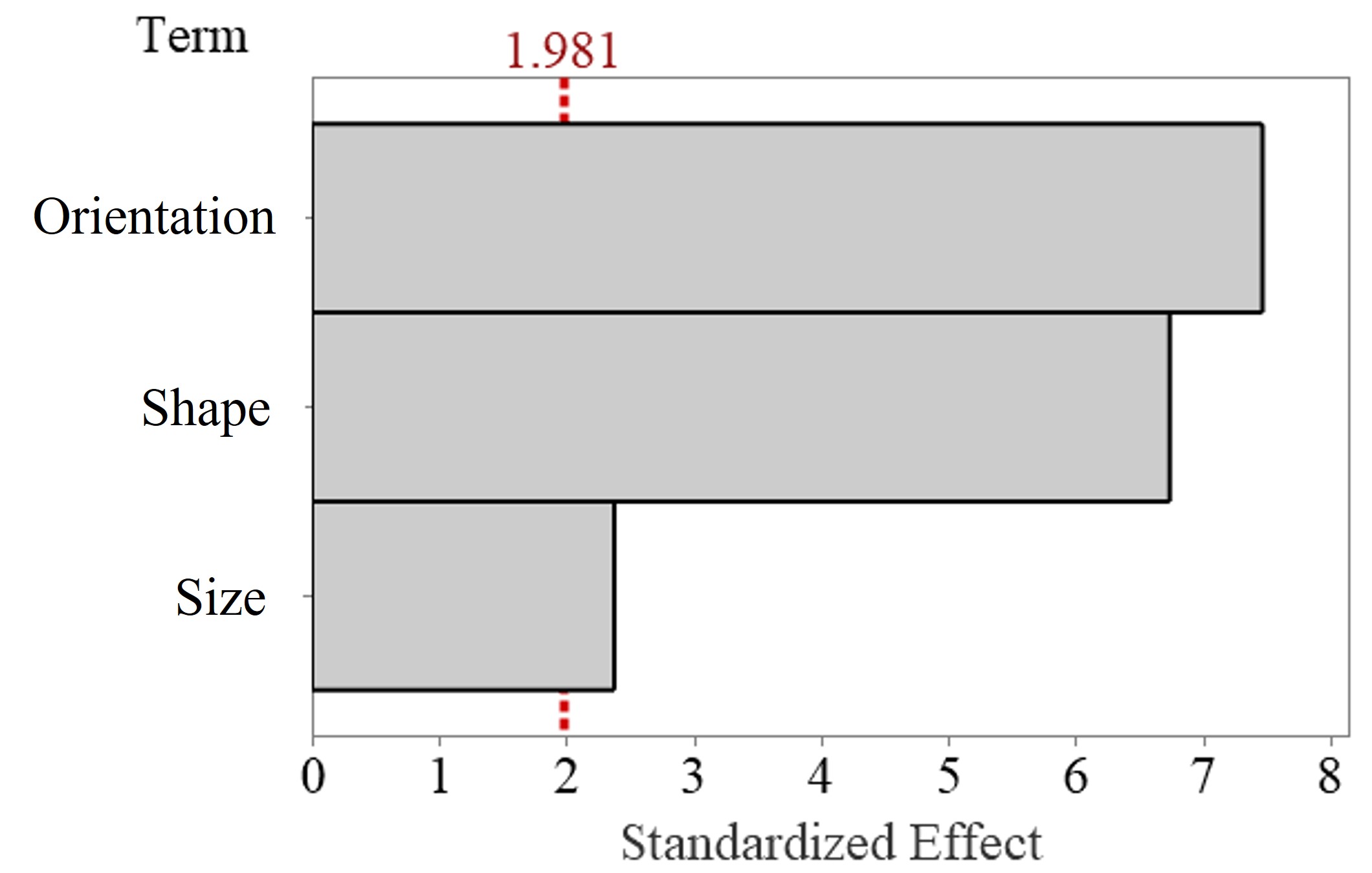}
	\centering
	\caption{Pareto chart of the standardized effects for backscatter amplitude predicted at a 95\% confidence interval. The red line (1.981) indicates that factors exceeding this are deemed to have a statistically significant impact.}
	\label{fig:Pareto_BS}
\end{figure}

Next, we evaluate the effects of different macrozones characteristics on backscatter amplitude. From the Pareto chart in Figure \ref{fig:Pareto_BS}, all three macrozone characteristics are deemed to be significant in generating backscatter signals, with orientation being the most significant, followed by macrozone shape and size.

\begin{figure}[h!]
	\includegraphics[width=130mm]{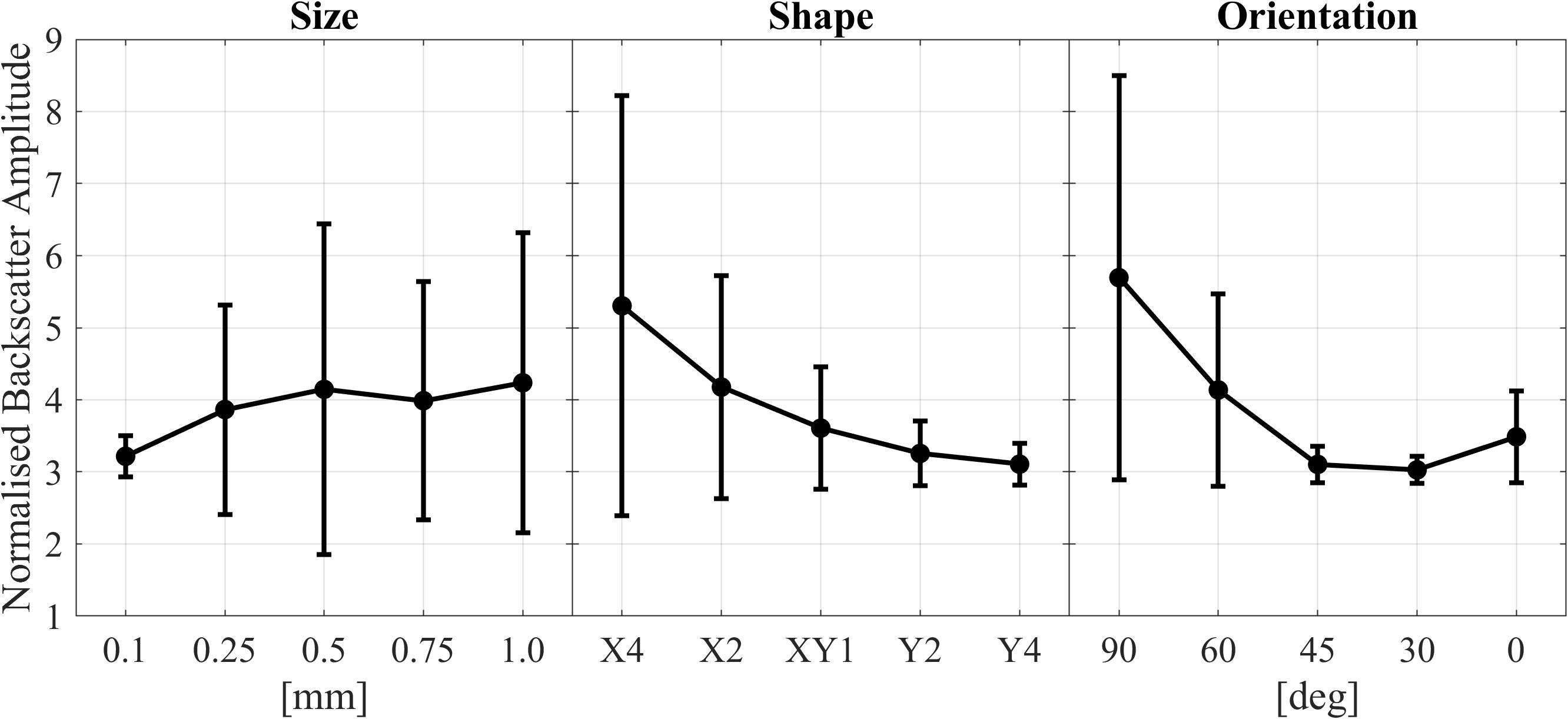}
	\centering
	\caption{The main effects plot for backscatter highlights the correlation between ultrasound backscatter and the three macrozone characteristics: (a) size, (b) shape, and (c) orientation.}
	\label{fig:Main_Effects_BS}
\end{figure}

Following up with the main effects plot in Figure \ref{fig:Main_Effects_BS}, there is a general positive correlation between backscatter amplitude and macrozone size. Contrastingly, backscatter amplitude is inversely related to macrozone elongation, whereby macrozone length increases along the wave propagation direction. Lastly, backscatter amplitude is highest when the macrozone C-axis orientation is parallel to the wave propagation direction as velocity and impedance mismatch is the greatest.

\begin{figure}[h!]
	\includegraphics[width=150mm]{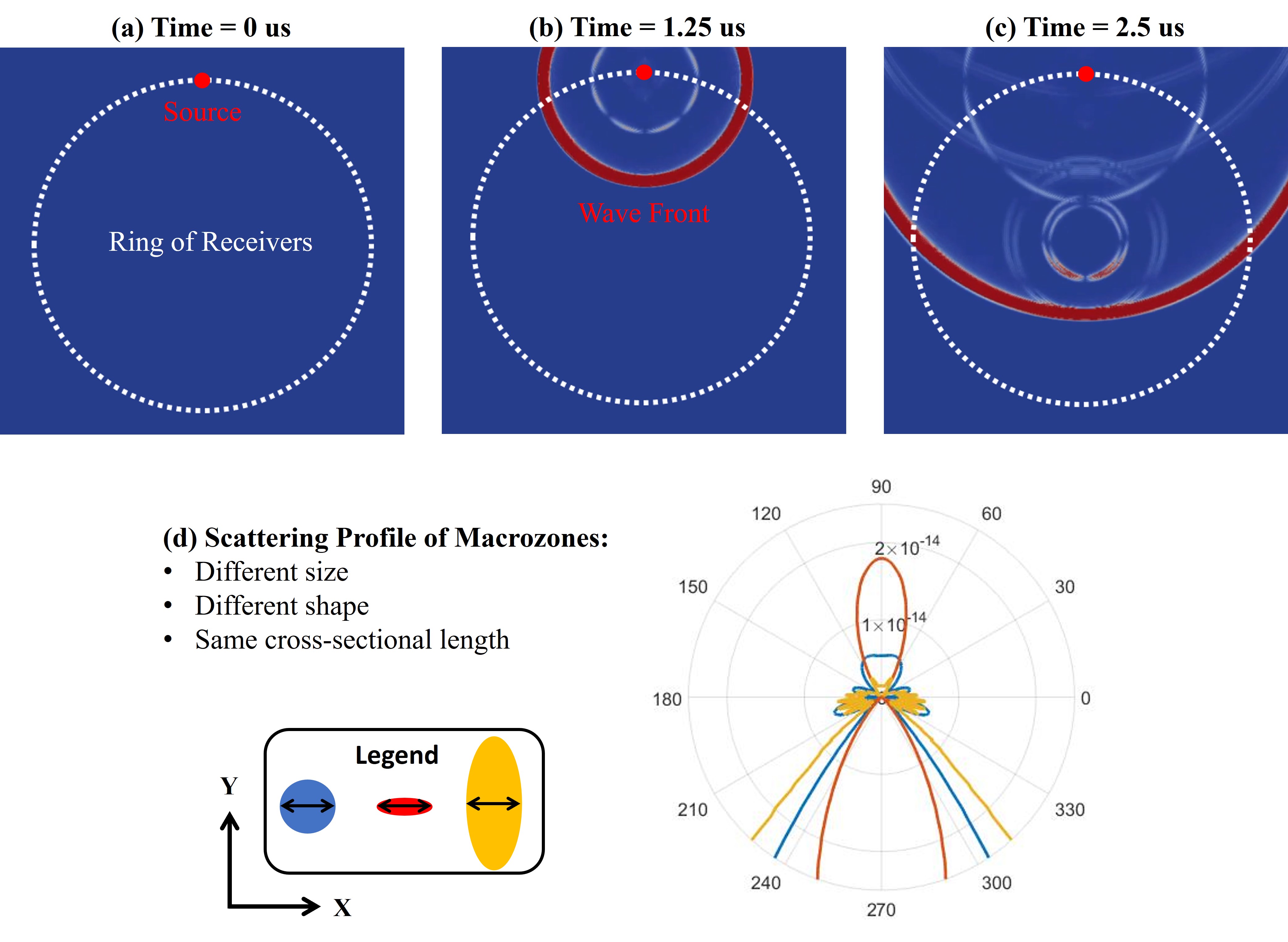}
	\centering
	\caption{(a) The model setup contains a ring of receivers represented by the white-dotted line with a source located at the top. (b) A longitudinal wave is excited which propagates downwards. (c) The propagated wave interacts with the macrozone at the centre which generates scattered signals in all directions. (d) The scattered profile generated with the wave incident from the 90-degree direction. The colored lines depict three macrozones with the same cross-sectional length but different sizes and shapes.}
	\label{fig:BS_FE_Model_Setup}
\end{figure}

This backscatter variation due to macrozone size and shape can be further explained using a scattering profile. This is generated with FE simulation as illustrated in Figure \ref{fig:BS_FE_Model_Setup}. The setup is shown in (a) which consists of a circular ring of receivers represented by the white-dotted line with a longitudinal excitation source located at the top. In (b) at time = 1.25 $\mu s$, the longitudinal wave is excited and propagates downwards. In (c) at time = 2.5 $\mu s$, the wave interacts with the macrozone located at the centre which generates longitudinal and mode-converted shear scattered signals in all directions which are captured by the ring of receivers. The longitudinal scattered signals are Fourier-Transformed to obtain the scattering amplitude at the excitation Frequency (10 MHz).

The scattering profiles generated for three different macrozones are illustrated in Figure \ref{fig:BS_FE_Model_Setup} (d), with the incident wave from the 90 degree direction. The macrozones are shown in the legend and contain different sizes and shapes but with the same cross-sectional length. Their scattering profiles are represented by the respective colored lines and we are only interested in the backscatter amplitude which is directed along the 90 degree angle. Even though the X-elongated macrozone (red) is the smallest, it generated the largest backscattered amplitude compared to the other two macrozones because of its preferential pancake shape. Hence, this demonstrates why macrozone shape has the greatest impact on backscatter amplitude, and a larger surface area normal to the wave propagation direction promotes stronger backscsattering signals.

\begin{figure}[h!]
	\includegraphics[width=140mm]{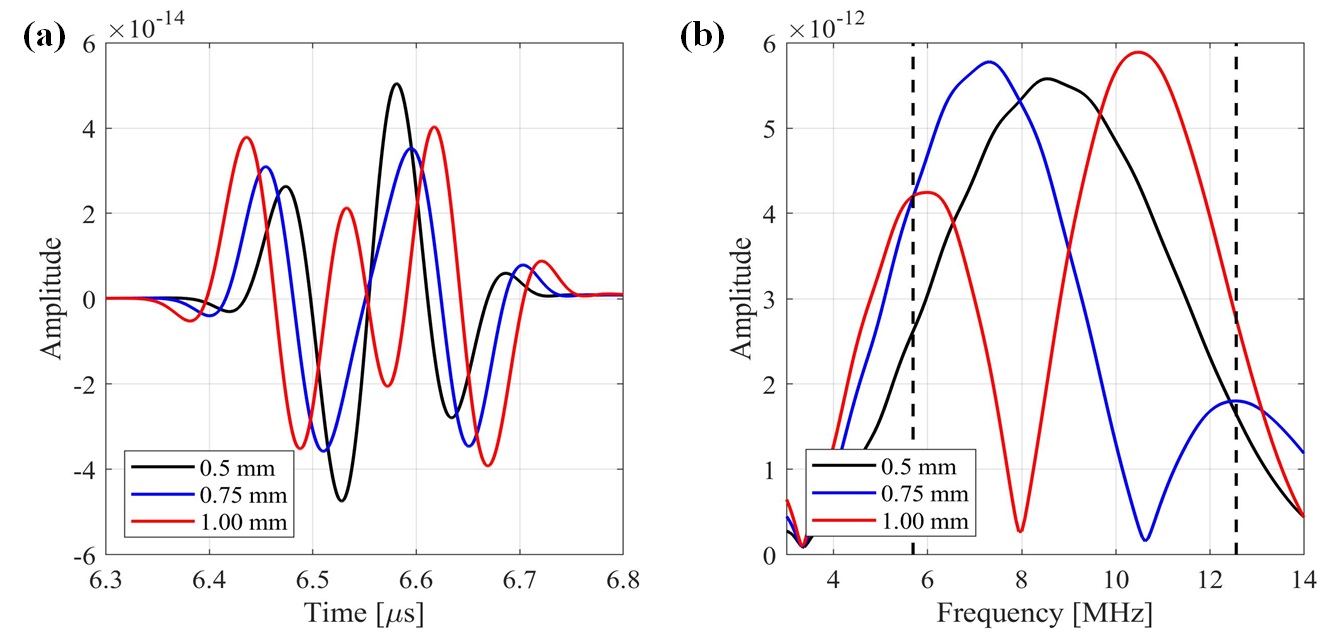}
	\centering
	\caption{(a) Time domain backscatter signals and (b) the respective frequency spectrum of single-macrozone models with varying sizes.}
	\label{fig:BS_Resonance}
\end{figure}

    In Figure \ref{fig:Main_Effects_BS}, we observe a dip in the mean backscatter profile with respect to different macrozone sizes at 0.75 mm diameter. The dip is a result of the macrozones having a dimension that is half the longitudinal wavelength along the wave propagation direction. This leads to a destructive interference between the backscattered signal generated from the top and bottom boundary of the macrozone. This effect is demonstrated using single-macrozone models as illustrated in Figure \ref{fig:BS_Resonance} which shows the backscatter (a) time-domain signal and (b) frequency response of three pancake (X4) macrozones with varying sizes. From both figures, the amplitude response from the 0.75 mm pancake macrozone is the smallest as its dimension along the wave propagation direction is closest to half the longitudinal wavelength, resulting in a distorted time domain signal and reduced backscatter intensity in the frequency spectrum. A similar phenomenon was also observed by Liu et al. \cite{Liu2019} whereby the backscatter amplitude from singular inclusions were modelled using FE and analytical solutions based on the Born approximations \cite{Rose1992}. This also highlights the potential of being able to use backscattering amplitude and the Born approximation to quantify macrozone dimensions.

\subsection{Velocity}
\begin{figure}[h!]
	\includegraphics[width=80mm]{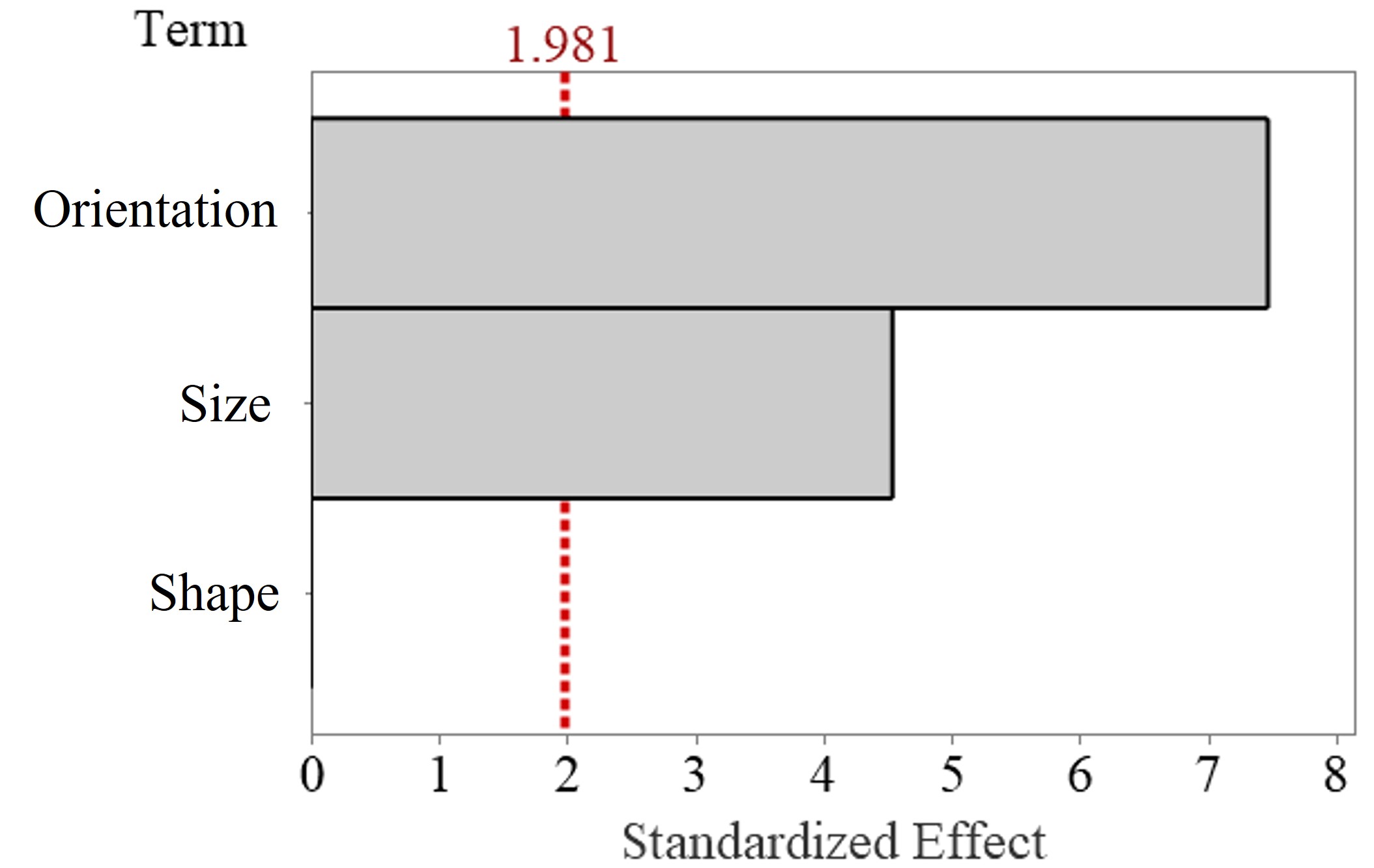}
	\centering
	\caption{Parteo chart of the standardized effects for velocity predicted at a 95\% confidence interval. The red line (1.981) indicates that factors exceeding this are deemed to have a statistically significant impact.}
	\label{fig:Pareto_Vel}
\end{figure}

Lastly, we will discuss the effects of different macrozone characteristics on ultrasound velocity. From the Pareto chart in Figure \ref{fig:Pareto_Vel}, macrozone crystal orientation has the largest impact, followed by size, and lastly the shape of the macrozone that is deemed to be statistically insignificant.

\begin{figure}[h!]
	\includegraphics[width=130mm]{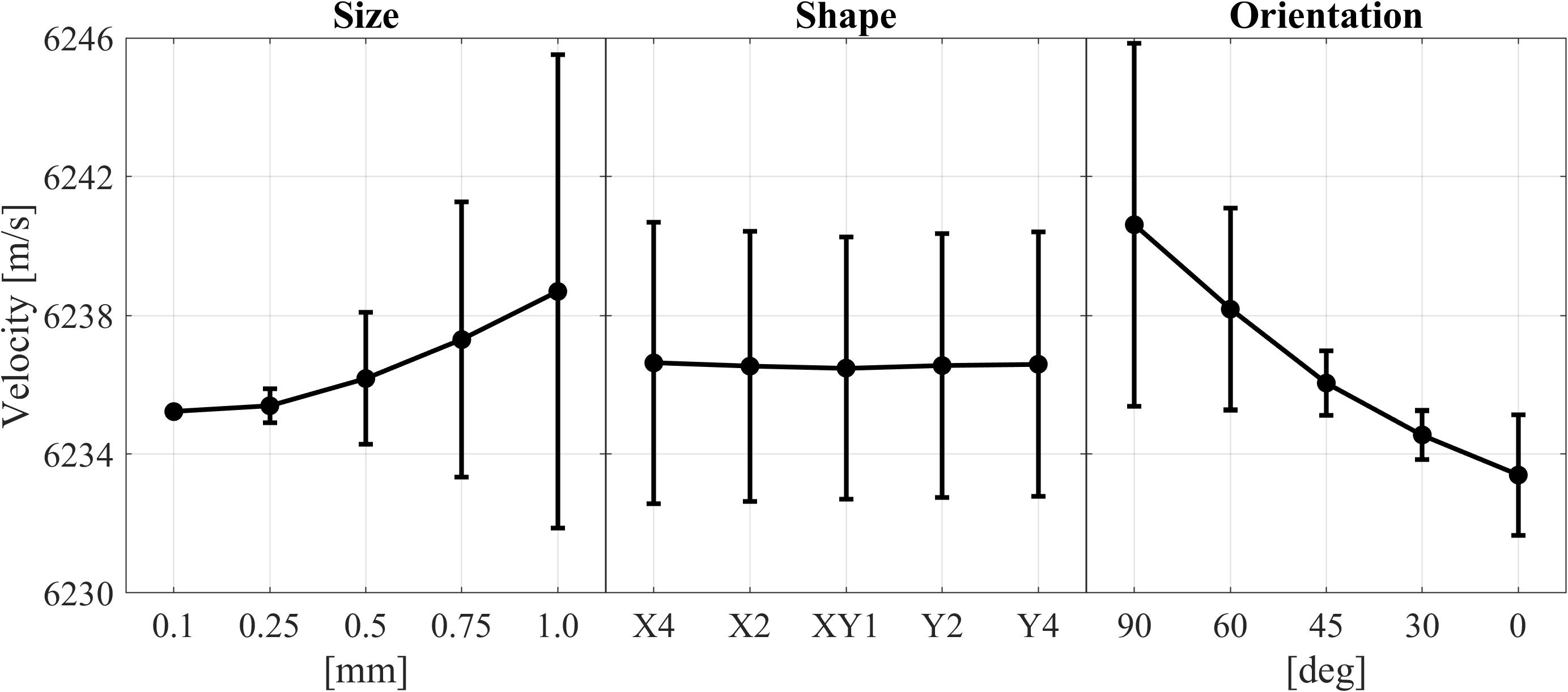}
	\centering
	\caption{The main effects plot for velocity highlights the correlation between ultrasound velocity and the three macrozone characteristics: (a) size, (b) shape, and (c) orientation.}
	\label{fig:Main_Effects_Velocity}
\end{figure}

Any variation in velocity is generally caused by the presence of texture (preferred crystal orientations) that is dependent on the area (2D) or volume (3D) of macrozones present \cite{Lan2014a}. As such, from the main effects plot in Figure \ref{fig:Main_Effects_Velocity}, ultrasound velocity increases with macrozone sizes, whereas any macrozone elongation does not affect the area of the macrozone and hence is insignificant towards ultrasound velocity. Lastly, velocity varies monotonically with crystal orientation, akin to the velocity profile in Figure \ref{fig:Velocity_Profile} where velocity is highest along the macrozone C-axis (oriented at 90 degree).

The actual variation in the mean velocity within the range of modelled parameters are minute, with only an 8 m/s difference for an orthogonal shift in the crystal orientation of the macrozones. Even though bulk texture is not considered in this study, it does indicate that small changes in velocity might not be easily detectable in Ti-64 samples with small volumes (3D) of MTR. 

\subsection{Discussion}
The wave behaviour in the Ti-64 alloy is expected to follow the conventional Rayleigh and Stochastic scattering mechanisms used to describe wave scattering from polycrystalline material \cite{Stanke1984a,Weaver1990,VanPamel2018a}. Hence, macrozones with sizes much smaller than the wavelength operate within the Rayleigh regime (kd <= 1), and macrozones with sizes comparable to the wavelength operate within the Stochastic regime (kd > 1). 

Within the Rayleigh regime, for the same size and crystal orientation, any shape variation in the macrozone has a limited effect on attenuation. However in the Stochastic regime, any elongation of the macrozones increases attenuation due to additional phase perturbation along the length of the macrozones and vice versa \cite{Huang2021}.

\begin{figure}[h!]
    \includegraphics[width=80mm]{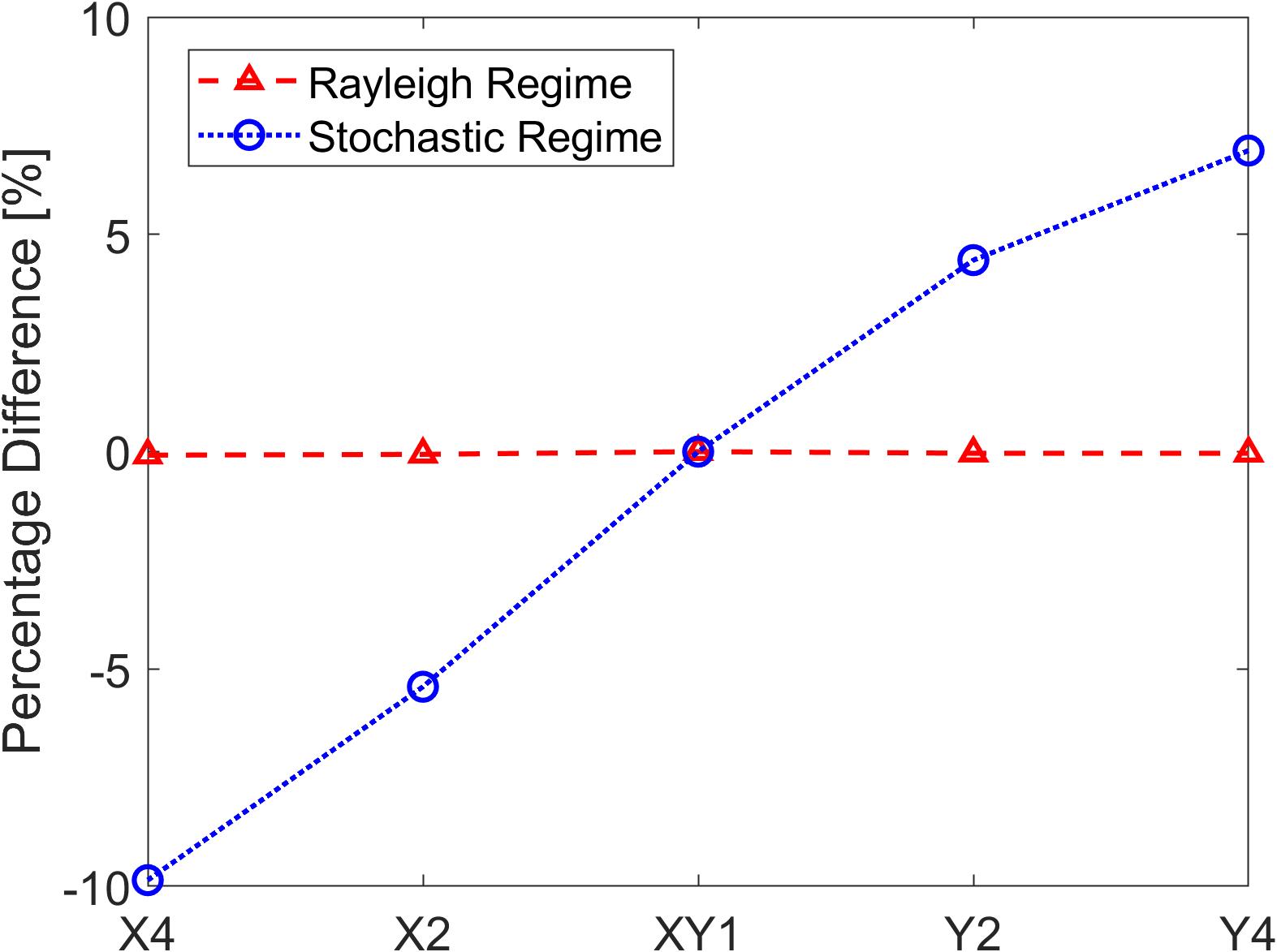}
	\centering
	\caption{Percentage difference in the simulated attenuation of different macrozone elongation ratios with respect to the equiaxed macrozone in the Rayleigh (kd $\approx$ 1) and Stochastic regimes (kd $\approx$ 10).}
	\label{fig:R_vs_S}
\end{figure}

 A comparison of the simulated attenuation for different macrozone shapes (with respect to the circular macrozone) in the Rayleigh and Stochastic regimes is shown in Figure \ref{fig:R_vs_S}. In the Rayleigh regime, the percentage difference calculated using the equiaxed case (XY1) as a reference holds steady at about zero, whereas in the Stochastic regime, attenuation increases monotonically with macrozone elongation along the wave direction.

Hence, to use ultrasound measurements to characterise macrozones, it is advisable to operate within the Stochastic regime, where the wavelengths are comparable to the morphologies of the macrozones, and have enhanced sensitivities to them than the smaller grains in the background. Macrozone sizes tpyically range from microns up to the millimetre scale \cite{Liu2021}, so the operating frequency should be about 5 to 15 MHz to remain within the Stochastic regime.

Looking back at the main effects plot for attenuation (Fig \ref{fig:Main_Effects_Att}), backscatter (Fig. \ref{fig:Main_Effects_BS}) and velocity (Fig \ref{fig:Main_Effects_Velocity}), the large variation in standard deviation indicates certain compounded  influence between the parameters. For example, in the attenuation plot with the largest macrozone size, a combination of X4 elongation with 45-degree crystal orientation results in a small attenuation of 0.8 Np/m, whereas a Y4 elongation with 90-degree crystal orientation leads to a higher attenuation of 4.2 Np/m. Hence, having certain macrozone characteristic combinations leads to a non-linear variation in the measured attenuation. The lack of independence of the parameters imposes some limitations on the possibility of drawing rigorous conclusions from the Pareto charts. Nevertheless, this approach is effective in providing key insights to the behavioural trends.

\begin{figure}[h!]
	\includegraphics[width=160mm]{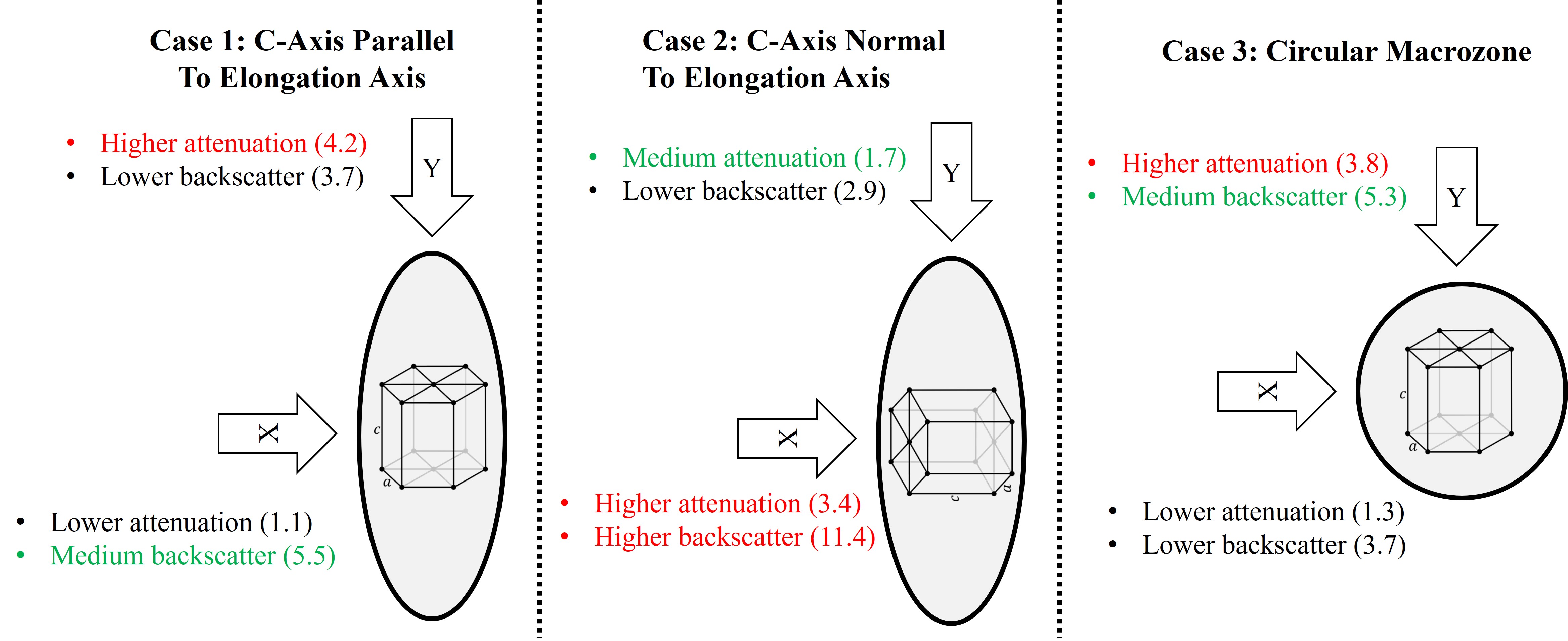}
	\centering
	\caption{The ultrasound attenuation (Np/m) and backscatter (arbitrary normalized amplitude) response with respect to three general cases of macrozones of 1 mm size in equivalent diameter are summarized in this figure. For case 1, the macrozone C-axis orientation is parallel to the elongation axis. In case 2, the macrozone C-axis orientation is normal to the elongation axis. In case 3, the circular macrozone has its C-axis orientation along the Y-direction.}
	\label{fig:DOE_General_Case}
\end{figure}

The wave-macrozone interactions can be summarised into three general cases as highlighted in Figure \ref{fig:DOE_General_Case}:

\begin{enumerate}
    \item In the first case, the C-axis of the macrozone is parallel to the grain elongation axis. High attenuation is expected along the Y-direction due to the preferred C-axis orientation leading to stronger wave distortion, and the elongated grain increases the phase perturbations further. Medium backscatter amplitude is obtained along the X-direction mainly due to the larger pancake shape and the preferable surface profile which leads to more coherent backscatter amplitude signals.
    
    \item In the second case, the C-axis of the macrozone is normal to the grain elongation axis. High attenuation is expected along the X-axis due to the preferred C-axis orientation as a result of stronger wave distortion. Even though the grain is elongated along the Y-axis, the DOE study has concluded that crystal orientation plays a more significant role as opposed to grain shape when it comes to ultrasonic attenuation measurements. As such, attenuation along the Y-axis is of a medium level, since the elongated length increases phase perturbations. As for backscattering amplitude, it will be higher along the X-axis due to the preferable surface profile and C-axis orientation.
    
    \item In the third case, the macrozone is spherical in shape. Since the macrozone is symmetrical in shape, the direction of higher attenuation and backscatter follows the C-axis orientation of the macrozone - which in this case is along the Y-axis - since there is a larger velocity mismatch between the macrozone and the background in this propagation direction.
\end{enumerate}

These cases demonstrate how different macrozone-parameter combinations give rise to varying ultrasound responses in orthogonal directions. Cases   1 and 2 both have the same elongated macrozones but due to them having a different C-axis orientation, the attenuation and backscatter responses vary significantly. It is crucial to note that the macrozones are normally generated through deformation processes such as rolling, and the preferred C-axis orientations usually follow the deformation and grain elongation direction \cite{Lunt2014c}. As such, cases 1 and 3 are more common than case 2. Nevertheless, these various 2D cases serve as a reference for experimental measurements and can provide some qualitative insights toward the general 3D macrozone configuration.

The spatial density of the macrozones is defined arbitrarily as the goal was to include multiple macrozones to emulate realistic samples. Even though the density of the macrozones are not the goal of this study, it will influence the result. Having less-dense macrozones will generally result in lower attenuation and backscatter, whereas velocity will depend on the overall bulk texture. However, there is also a maximum point whereby further increment in macrozone density and assuming that they overlap will result in the medium being akin to a single crystal and hence lead to reduced attenuation and backscatter. 

\section{Experimental Confirmation}
\begin{figure}[h!]
	\includegraphics[width=160mm]{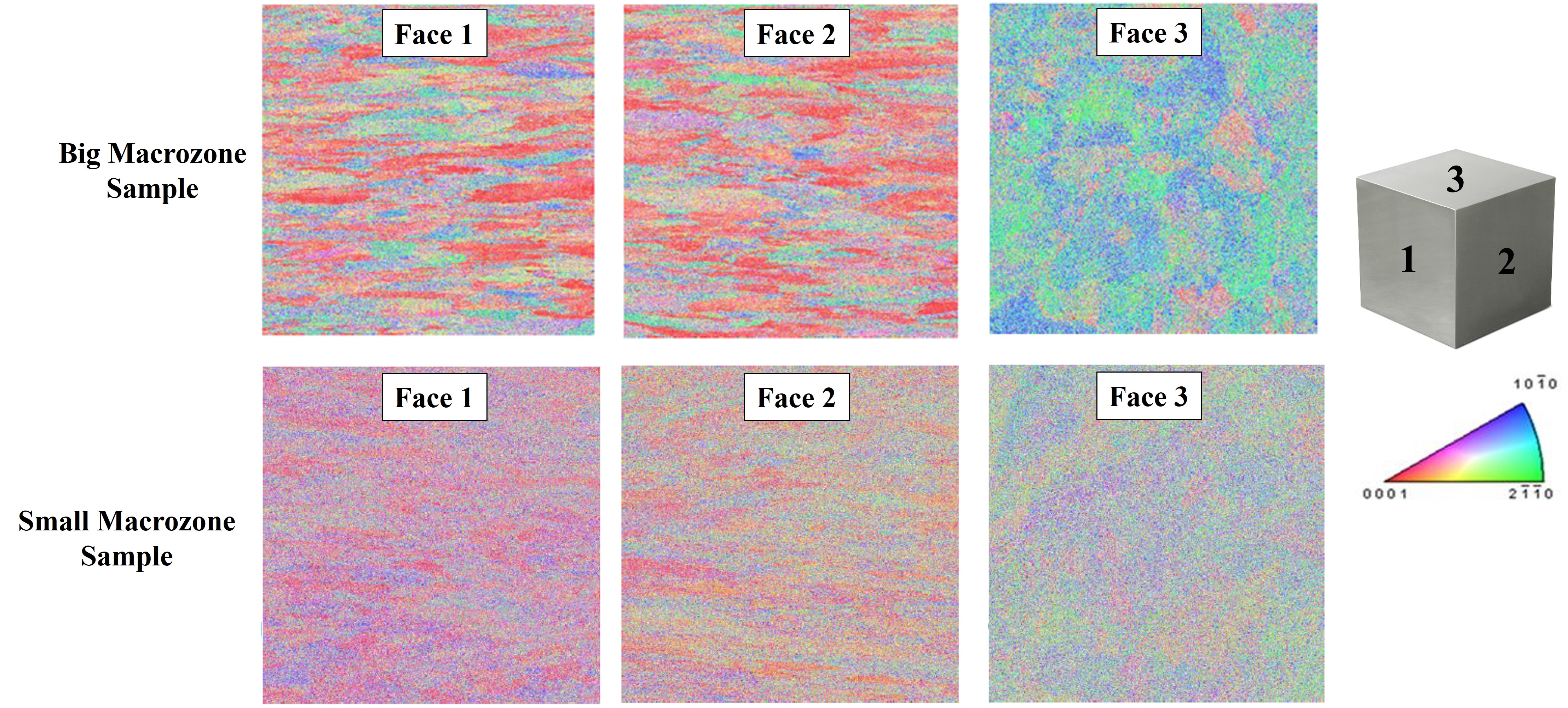}
	\centering
	\caption{The EBSD images of all 3 faces of the big- and small-MTR Ti-64 cubic sample are shown with a dimension of 10 x 10 mm. The color code indicates the orientation of the grains as depicted by the Inverse Pole Figure.}
	\label{fig:EBSD_Ti64}
\end{figure}

Experiments were conducted on sample cases in order to validate the modelling results. The tests are conducted on two cubic Ti-64 samples provided by industrial partner IHI with a dimension of 45 mm. The samples are named big and small macrozone samples, as they underwent different heat treatment and deformation processes to introduce different levels of macrozones. The EBSD images of the samples are illustrated in Figure \ref{fig:EBSD_Ti64}. The EBSD images of all three faces of the samples are shown with the color code indicating the orientation of the grains. The cluster of reds in Faces 1 and 2 in both the big- and small-MTR samples indicate the presence of elongated macrozones, with the big-MTR sample having condensed and visually larger macrozones present. 

\begin{figure}[h!]
        \includegraphics[width=90mm]{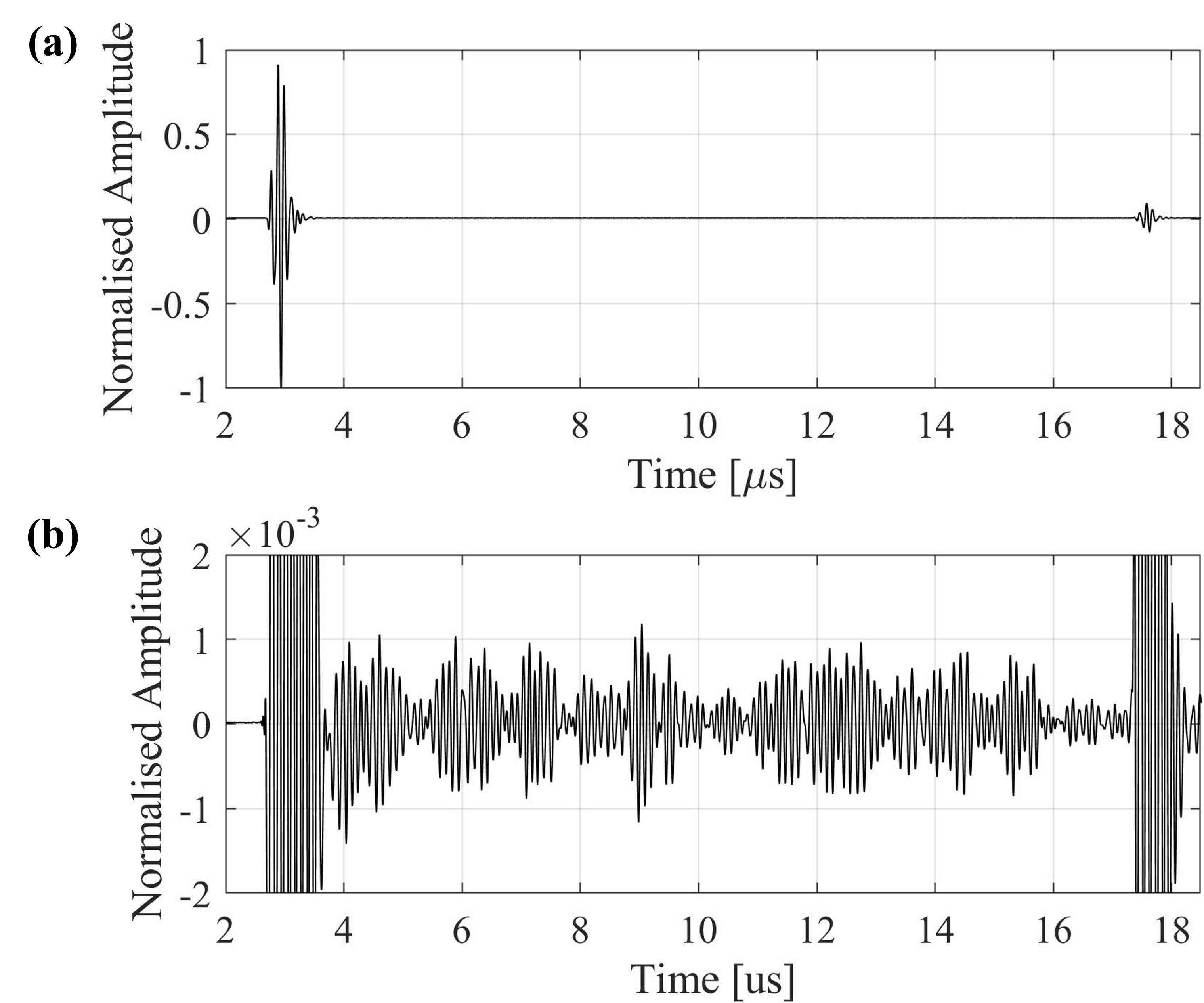}
	\centering
	\caption{Experimental Signals of (a) front and back wall reflections, and (b) backscatter signals with front and back wall saturated measured from the Big-MTR sample.}
	\label{fig:Experimental_Signal}
\end{figure}

The samples are measured on all three surfaces using water-immersion pulse-echo configuration and a 10 MHz planar 6.35 mm diameter transducer with a stand-off distance of 25 mm and a step size of 0.2 mm. The spot size of the transducer beam on the sample surface was approximated using a Huygens’ model to be about 4 mm. The Peak NDT LT2 ultrasonic inspection system is used for this experiment. Signals are acquired at a sampling rate of 200 MHz with 128 averages to reduce electrical noise. A sample signal is shown in Figure \ref{fig:Experimental_Signal}, where the top figure shows a time-domain signal with the front and back wall that are used for attenuation and velocity measurements, and the bottom figure shows a saturated signal that captures the low-amplitude backscattered signals. Similar to the FE models, ultrasonic attenuation and velocity are measured and calculated with the equation \ref{att} \cite{Zeng2010} and the cross-correlation method \cite{Zhang2008} respectively. This is then averaged over 120 x 120 spatial positions. As for the backscattered signals, the root-mean-square of the backscattered signals along the 120 x 120 positions are obtained in the frequency domain to attain the backscatter spectral response \cite{Margetan1993}.

\begin{figure}[h!]
    \includegraphics[width=160mm]{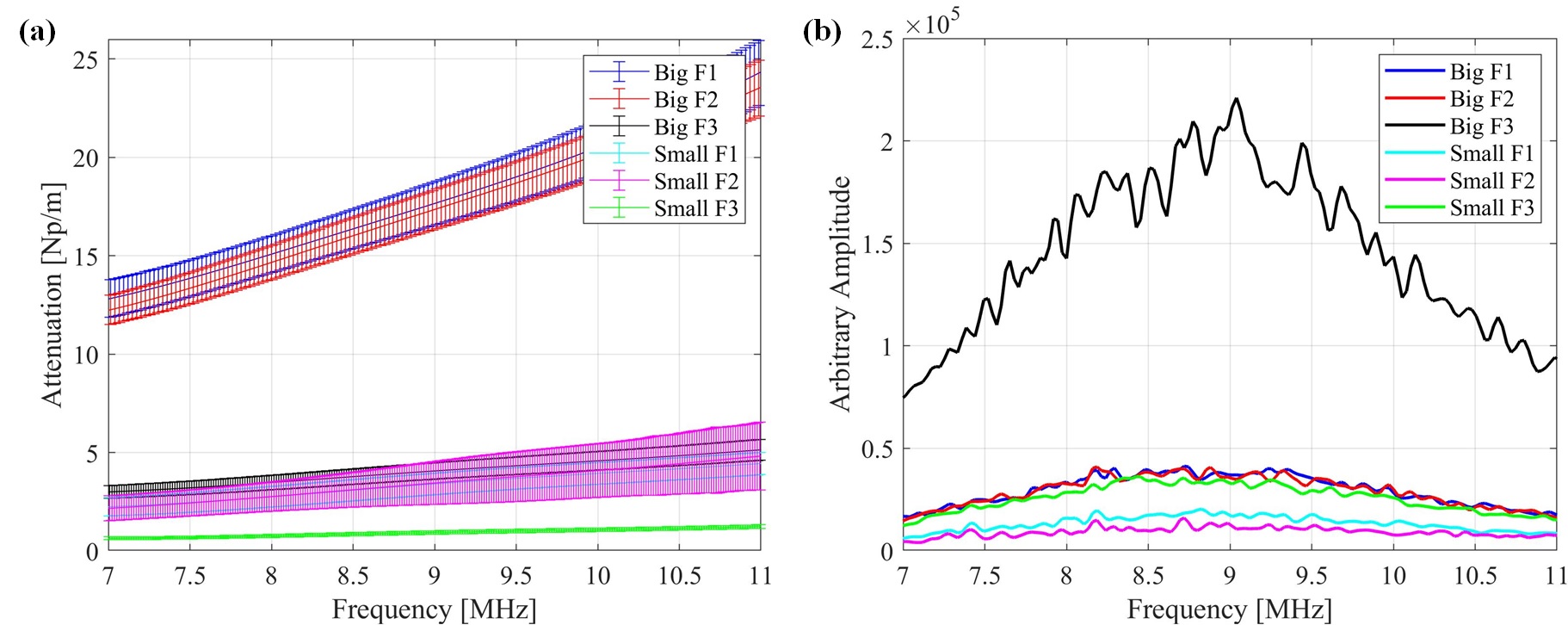}
	\centering
	\caption{Experimentally measured (a) average attenuation and (b) backscatter root-mean-squared for both big and small MTR samples along all 3 orthogonal faces.}
	\label{fig:Experimental_BS_Att}
\end{figure}

The results are shown and compared in Figure \ref{fig:Experimental_BS_Att} and Table \ref{tab:Results}. In the big-MTR sample, backscatter amplitude is much higher along face 3, whereas attenuation is much higher along faces 1 and 2. Higher attenuation is obtained along faces 1 and 2 due to the preferred C-axis orientations which leads to a stronger beam-distortion effect, and a longer mean-grain length which further contributes to the phase perturbation. Along face 3, backscatter amplitude is much stronger due to a larger macrozone surface area with respect to the wave propagation direction. These opposing behaviours across orthogonal scanning faces were also seen in \cite{Keller2005,Lobkis2012a}. The standard deviation in the attenuation profile represent the variation in attenuation measurements as a result of the macrozones having a random spatial distribution within the sample. As such, the wave encounters different macrozone sizes and density at different spatial locations and hence give rise to varying attenuation values.

The measured velocities for both samples are listed in Table \ref{tab:Results}, with velocities being higher along faces 1 and 2 in both samples. The velocity variation in the big MTR sample is about 112 m/s, whereas the variation in the small MTR sample is lower at about 56 m/s. The lower velocity variation in the small MTR sample also indicates the presence of smaller or less dense macrozones and therefore a weaker texture.

This trend is also consistent in the small-MTR sample, albeit with a smaller difference in the measured values. The smaller values in the small-MTR sample indicates that the macrozones in the sample are less pronounced than in the big-MTR sample. This can also be observed in the EBSD data shown in Figure \ref{fig:EBSD_Ti64} where the grain clusters are smaller in general. This finding was also corroborated by the parametric study generated using FE models for attenuation (Figure \ref{fig:Main_Effects_Att}) and backscatter (Figure \ref{fig:Main_Effects_BS}), and is also reported in an experimental study in \cite{Bhattacharjee2011}.

\begin{figure}[h!]
    \includegraphics[width=130mm]{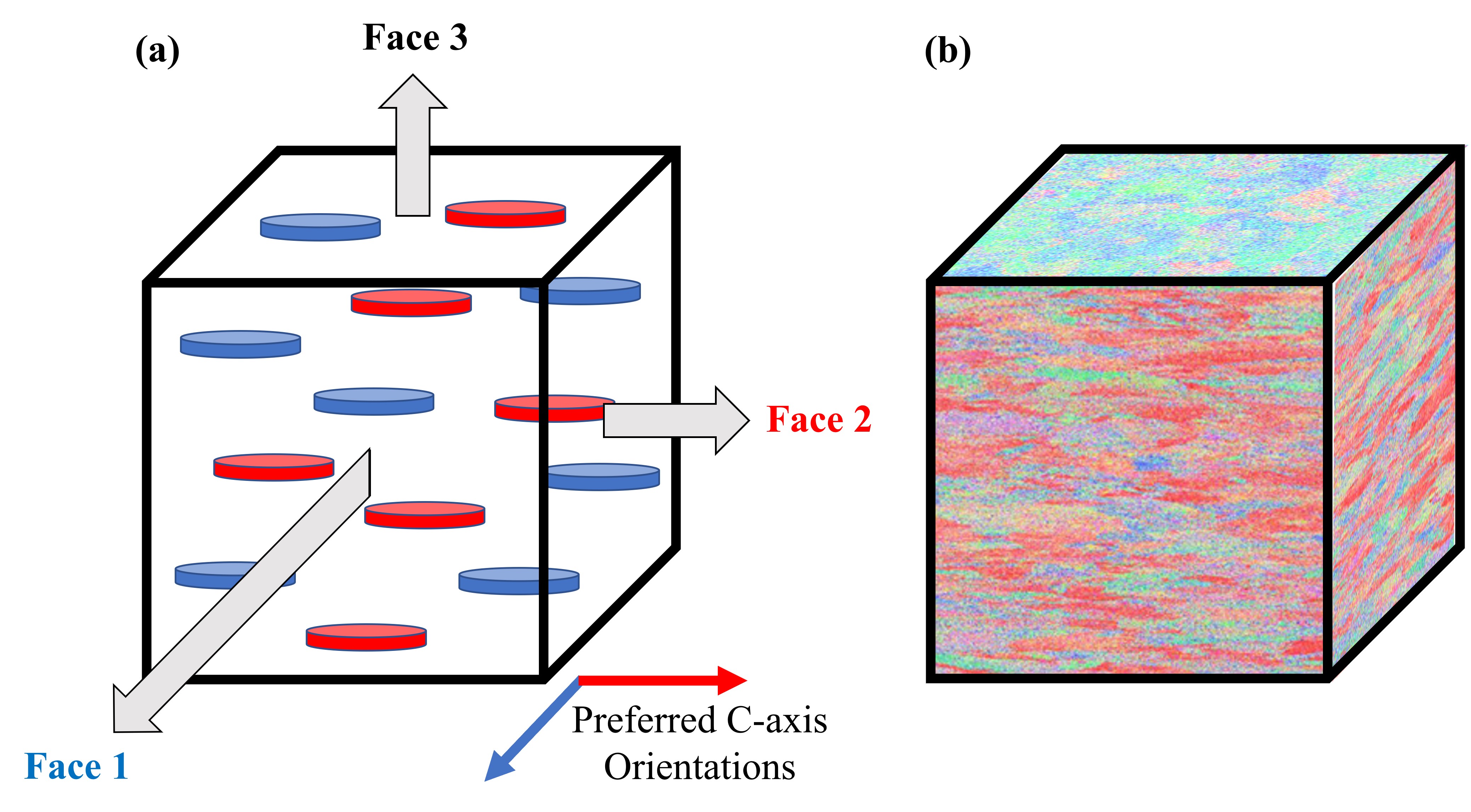}
	\centering
	\caption{(a) The macrozones in the big MTR sample are visualised as pancake-shaped with two preferred C-axis orientations along faces 1 and 2. (b) EBSD results along all three faces whereby the color red represents macrozones having their C-axes pointing out of the surface.}
	\label{fig:Pancake}
\end{figure}

Next, we relate the experimental results with the findings obtained from the FE models to evaluate the characteristics of the macrozones present in our samples by comparing the ultrasound responses in Table 3. From section 3 with reference to Fig 17 case 1, it was deemed that attenuation is generally higher along C-axis and elongation directions, whereas backscatter is generally higher across elongation directions. Likewise, from the small- and big-MTR results measured at 10 MHz, attenuation and velocity is higher along faces 1 and 2, indicating the preferred C-axes and macrozone elongation are oriented along these faces. Whereas high backscatter amplitude along face 3 indicates that the macrozone has a wider surface profile. This enables us to visualise the macrozones as having a pancake shape, and this is corroborated with the EBSD results along the three faces, as shown in  Figure \ref{fig:Pancake}. Similar findings were also observed in \cite{Margetan2002,Margetan2005}, with high attenuation and wave distortion along the elongated and C-axis directions.

\begin{table}[hbt]
\centering
\caption{Attenuation, backscatter ratio, and velocity measured from FE model (case 1 in Fig. 17) and experiments with the small- and big-MTR samples.}
\begin{tabular}{ccccc}
\cmidrule{3-5} & & Attenuation & Backscatter & Velocity\\
\midrule
FE Model & Elongated (E) & 4.2 & 1.5 & 6250 \\
 & Shortened (S) & 1.1 &  & 6230 \\
 \midrule
 & Face 1 (E) & 3.5 & & 6229 \\
Small MTR Sample & Face 2 (E) & 3.5 & 2 & 6216 \\
& Face 3(S) & 0.9 &  & 6166\\
  \midrule
 & Face 1 (E) & 20.5 & & 6254 \\
Big MTR Sample & Face 2 (E) & 20.1 & 5 & 6256 \\
& Face 3(S) & 4.6 &  & 6143\\
\bottomrule
\end{tabular}
\label{tab:Results}
\end{table}

 In general ultrasound measurements, we expect a signal that has high attenuation to have high backscatter (or vice versa) since any form of scattering within the medium should reduce the overall energy of the wave. Nevertheless, in the case of the big-MTR sample where the wave propagates along faces 1 and 2, we get low backscatter and high attenuation and the energy of the wave appears to be missing. As explained, the low backscatter amplitude is a result of the needle-like macrozone surface profile which does not generate significant backscatter signals. The forward-propagating wave is however strongly affected by the preferred C-axis orientation which distorts the wave front. Due to this phase perturbation, the wave captured by the receiving nodes undergoes destructive interference and has a lower amplitude - similar beam distortion effects as a result of the macrozones were reported in \cite{Margetan2005}. The total energy of the wave, on the other hand, is still preserved - this was evaluated using FE models at 10 MHz which is the same as the experimental operating frequency.

\section{Conclusion}
From this study, we are able to generate idealised 2D FE models that are qualitatively representative of actual complex microstructures containing macrozones at grain-scale and have validated the findings experimentally.

Next, we highlighted important physical understanding of the wave-macrozone interactions and the trends observed with different types of macrozones. It is important to note that the general relationships observed here are also valid for general inclusions of varying sizes and shapes. The systematic study using DOE could also be expanded, if it is useful in other contexts, to include other types of features, such as having diagonally aligned macrozones and multiple preferred C-axis orientations.

Lastly, we demonstrated how the experimental results can be used for initial qualitative characterisation of the bulk macrozones within the sample. Based on orthogonal experimental scans and identifying trends in the results, we can determine the overall shape and textured directions of the macrozones.
\\\\\\
\bibliographystyle{unsrt}  
\bibliography{references}

\end{document}